\documentclass[twoside,twocolumn,9pt]{article}
\usepackage{extsizes}
\usepackage[super,sort&compress,comma]{natbib} 
\usepackage[left=1.5cm, right=1.5cm, top=1.785cm, bottom=2.0cm]{geometry}
\usepackage{balance}
\usepackage{mathptmx}
\usepackage{sectsty}
\usepackage{graphicx} 
\usepackage{lastpage}
\usepackage[format=plain,justification=justified,singlelinecheck=false,font={stretch=1.125,small,sf},labelfont=bf,labelsep=space]{caption}
\usepackage{float}
\usepackage{fancyhdr}
\usepackage{fnpos}
\usepackage[english]{babel}
\addto{\captionsenglish}{%
  
}
\usepackage{array}
\usepackage{droidsans}
\usepackage{charter}
\usepackage[T1]{fontenc}
\usepackage[usenames,dvipsnames]{xcolor}
\usepackage{setspace}
\usepackage[compact]{titlesec}
\usepackage{hyperref}

\usepackage{epstopdf}

\usepackage{bm}
\usepackage{amsmath, cases}
\usepackage{amssymb}
\usepackage{mathtools, cases}   

\definecolor{cream}{RGB}{222,217,201}

\begin{document}

\pagestyle{fancy}
\thispagestyle{plain}
\fancypagestyle{plain}{
\renewcommand{\headrulewidth}{0pt}
}

\makeFNbottom
\makeatletter
\renewcommand\LARGE{\@setfontsize\LARGE{15pt}{17}}
\renewcommand\Large{\@setfontsize\Large{12pt}{14}}
\renewcommand\large{\@setfontsize\large{10pt}{12}}
\renewcommand\footnotesize{\@setfontsize\footnotesize{7pt}{10}}
\makeatother

\renewcommand{\thefootnote}{\fnsymbol{footnote}}
\renewcommand\footnoterule{\vspace*{1pt}%
\color{cream}\hrule width 3.5in height 0.4pt \color{black}\vspace*{5pt}} 
\setcounter{secnumdepth}{5}

\makeatletter 
\renewcommand\@biblabel[1]{#1}            
\renewcommand\@makefntext[1]%
{\noindent\makebox[0pt][r]{\@thefnmark\,}#1}
\makeatother 
\renewcommand{\figurename}{\small{Fig.}~}
\sectionfont{\sffamily\Large}
\subsectionfont{\normalsize}
\subsubsectionfont{\bf}
\setstretch{1.125} 
\setlength{\skip\footins}{0.8cm}
\setlength{\footnotesep}{0.25cm}
\setlength{\jot}{10pt}
\titlespacing*{\section}{0pt}{4pt}{4pt}
\titlespacing*{\subsection}{0pt}{15pt}{1pt}

\fancyfoot{}
\fancyfoot[LO,RE]{\vspace{-7.1pt}\includegraphics[height=9pt]{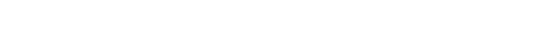}}
\fancyfoot[CO]{\vspace{-7.1pt}\hspace{13.2cm}\includegraphics{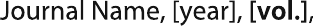}}
\fancyfoot[CE]{\vspace{-7.2pt}\hspace{-14.2cm}\includegraphics{RF}}
\fancyfoot[RO]{\footnotesize{\sffamily{1--\pageref{LastPage} ~\textbar  \hspace{2pt}\thepage}}}
\fancyfoot[LE]{\footnotesize{\sffamily{\thepage~\textbar\hspace{3.45cm} 1--\pageref{LastPage}}}}
\fancyhead{}
\renewcommand{\headrulewidth}{0pt} 
\renewcommand{\footrulewidth}{0pt}
\setlength{\arrayrulewidth}{1pt}
\setlength{\columnsep}{6.5mm}
\setlength\bibsep{1pt}

\makeatletter 
\newlength{\figrulesep} 
\setlength{\figrulesep}{0.5\textfloatsep} 

\newcommand{\topfigrule}{\vspace*{-1pt}%
\noindent{\color{cream}\rule[-\figrulesep]{\columnwidth}{1.5pt}} }

\newcommand{\botfigrule}{\vspace*{-2pt}%
\noindent{\color{cream}\rule[\figrulesep]{\columnwidth}{1.5pt}} }

\newcommand{\dblfigrule}{\vspace*{-1pt}%
\noindent{\color{cream}\rule[-\figrulesep]{\textwidth}{1.5pt}} }

\makeatother

\twocolumn[
  \begin{@twocolumnfalse}
{\includegraphics[height=30pt]{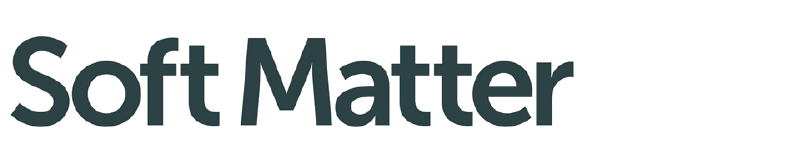}\hfill\raisebox{0pt}[0pt][0pt]{\includegraphics[height=55pt]{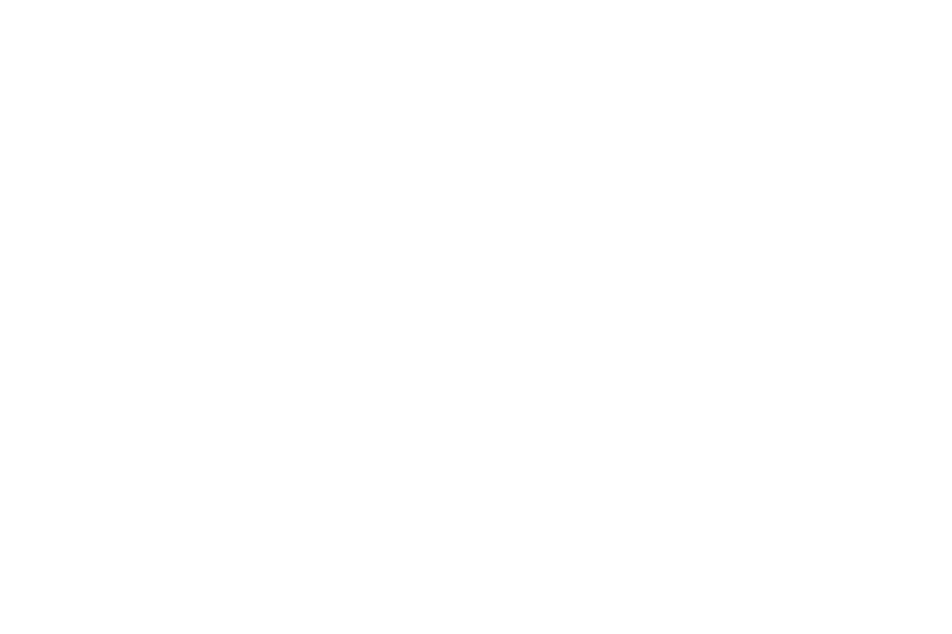}}\\[1ex]
\includegraphics[width=18.5cm]{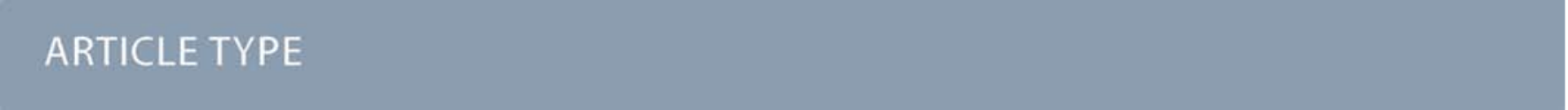}}\par
\vspace{1em}
\sffamily
\begin{tabular}{m{4.5cm} p{13.5cm} }

\includegraphics{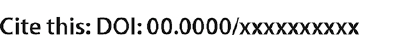} & \noindent\LARGE{\textbf{Glassy dynamics of sticky hard spheres beyond the mode-coupling regime
}} \\
\vspace{0.3cm} & \vspace{0.3cm} \\

 & \noindent\large{Chengjie Luo \textit{$^{a}$} and Liesbeth~M.~C. Janssen \textit{$^{a}$} } \\

\includegraphics{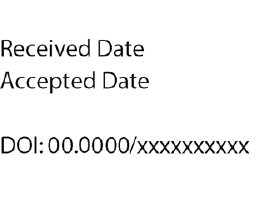} & \noindent\normalsize{Sticky hard spheres, i.e., hard particles decorated with a short-ranged attractive interaction potential, constitute a relatively simple model with highly non-trivial glassy dynamics. The mode-coupling theory of the glass transition (MCT) offers a qualitative account of the complex reentrant dynamics of sticky hard spheres, but the predicted glass transition point is notoriously underestimated. Here we apply an improved first-principles-based theory, referred to as generalized mode-coupling theory (GMCT), to sticky hard spheres. This theoretical framework seeks to go beyond MCT by hierarchically expanding the dynamics in higher-order density correlation functions -- an approach that may become exact if sufficiently many correlations are taken into account.  We predict the phase diagrams from the first few levels of the GMCT hierarchy and the dynamics-related critical exponents, all of which are much closer to the empirical observations than MCT. Notably, the prominent reentrant glassy dynamics, the glass-glass transition, and the higher-order bifurcation singularity classes ($A_3$ and $A_4$) of sticky hard spheres are found to be preserved within GMCT at arbitrary order. Moreover, we demonstrate that when the hierarchical order of GMCT increases, the effect of the short-ranged attractive interactions becomes more evident in the dynamics. This implies that GMCT is more sensitive to subtle microstructural differences than MCT, and that the framework provides a promising first-principles approach to systematically go beyond the MCT regime.} \\

\end{tabular}

 \end{@twocolumnfalse} \vspace{0.6cm}

  ]

\renewcommand*\rmdefault{bch}\normalfont\upshape
\rmfamily
\section*{}
\vspace{-1cm}


\footnotetext{\textit{$^{a}$~Department of Applied Physics, Eindhoven University of Technology, Eindhoven, The Netherlands. E-mail: l.m.c.janssen@tue.nl}}




\section{Introduction}
The dynamics of supercooled liquids and the nature of the glass transition are still not fully understood from a theoretical perspective.\cite{debenedetti2001supercooled,berthier2011theoretical} It is empirically well established that supercooling of a liquid is accompanied by a spectacular slowdown in dynamics over many orders of magnitude; at the same time, however, the disordered microstructure of the material remains very close to that of a normal liquid.\cite{gotze2008complex} 
One of the major challenges is to explain this emergent slow relaxation dynamics of glass-forming liquids on the basis of the material's static structure. At present, there are many theories that seek to answer this question, ranging from e.g.\ geometric-frustration-based theories\cite{tarjus2005frustration} to spin-glass-inspired approaches such as Random First Order Transition theory.\cite{kirkpatrick2015colloquium} More recently, machine learning has also been used to search for the link between structure and dynamics in dense disordered systems. \cite{liu2019machine,ma2019heterogeneous,schoenholz2016structural,bapst2020unveiling} However, there is still no complete and formally exact theory of glassy dynamics that is founded entirely on first principles. 

Mode-coupling theory (MCT), developed in the 1980s by G\"{o}tze and co-workers,\cite{bengtzelius1984dynamics,leutheusser1984dynamical,gotze2008complex,reichman2005mode,janssen2018mode} is an approximate framework that is widely regarded as the only first-principles-based theory of the glass transition. 
Among the main successes of MCT is its qualitative prediction of the non-trivial dynamical slowdown -- as characterized by the intermediate scattering function $F(k,t)$ for a given wavenumber $k$ and time $t$ -- based on only small changes in the static structure factor $S(k)$. For supercooled liquids, MCT typically predicts a two-step decay of $F(k,t)$ via the so-called $\beta$- and $\alpha$-relaxation processes, respectively, with a long plateau emerging at intermediate time scales. 
This plateau, which becomes longer upon deeper supercooling,  is indicative of  solid-like behavior and can also be understood in terms of the cage effect. Moreover, MCT predicts several scaling laws for the intermediate scattering function in the $\beta$- and $\alpha$-relaxation regimes that are generally consistent with the behavior seen in experimental glass-forming systems. 
When the temperature further decreases to a certain value, MCT predicts a liquid-glass transition at which the long-time limit of the intermediate scattering function, also called the non-ergodicity parameter, jumps discontinuously from zero to a positive value. This ergodicity-breaking transition is also a successful prediction from MCT.\cite{reichman2005mode}

However, owing to its approximate nature, MCT is also known to become inaccurate or can even break down in certain cases. The main MCT approximation stems from the fact that the intermediate scattering function $F(k,t)$, which in fact is a two-point density correlation function, is determined by a memory function in which four-point density correlation functions are required as the leading terms. In standard MCT, these four-point density correlation functions are approximated by the product of two $F(k,t)$ functions at different wavenumbers.\cite{gotze2008complex} This approximation is generally not accurate enough and is even uncontrolled.\cite{janssen2018mode} Additionally, the four-point density correlation functions may contain information on non-trivial spatiotemporal density fluctuations related to dynamical heterogeneity, but these non-trivial correlations are inherently lost when applying the MCT factorization approximation.\cite{kob1997dynamical,glotzer2000time, lavcevic2003spatially, zhang2020revealing} Hence, a more accurate theory taking multi-point density correlators into account is needed.

Recently, a new theory called generalized mode-coupling theory (GMCT) has been developed based on the framework of MCT.\cite{szamel2003colloidal,wu2005high,janssen2015microscopic} Within this theory, the uncontrolled approximation for the four-point density correlation functions is avoided by developing an exact dynamical equation for the four-point density correlators themselves; these are governed by a new memory function that is dominated by six-point density correlations, which in turn are controlled by eight-point correlators, et cetera. One may then apply a generalized MCT closure approximation at an arbitrary level in this hierarchy, thus delaying the uncontrolled factorization to a later stage. 
In this way, more accurate predictions for the relaxation dynamics can be obtained by systematically including higher-level correlation functions, which in fact establishes a hierarchy of coupled integro-differential equations. We note that, similar to MCT, the only material-dependent input for the theory is still the static structure factor $S(k)$ for any given temperature and density.

GMCT has been well studied for hard-sphere systems with a hard-core repulsive interaction potential.\cite{janssen2015microscopic,luo2020generalized1, luo2020generalized2} It has been found that GMCT indeed systematically improves the predictions from MCT. More specifically, the location of the liquid-glass transition point, which is usually underestimated in MCT, improves in a seemingly convergent manner as higher-order density correlations are included in the the first few levels of the GMCT hierarchy.\cite{szamel2003colloidal,wu2005high,janssen2015microscopic, luo2020generalized1}  Moreover, 
the qualitative MCT scaling laws near the liquid-glass transition in the $\beta$- and $\alpha$-relaxation regimes are fully preserved in GMCT, but with quantitatively systematically improved exponents.\cite{luo2020generalized1,luo2020generalized2} 
These triumphs of GMCT make it a promising first-principles-based theory for the glass transition, although it must be noted that the computational cost of a numerical GMCT calculation heavily increases with the number of included hierarchical levels, both in terms of computing time and required computer memory.\cite{luo2020generalized1}

Despite the promising GMCT results for hard spheres, many real glass-forming materials are governed by more complex interaction potentials between the particles. Standard MCT has already been widely tested for different glass-forming materials, including model systems interacting through Lennard-Jones potentials,\cite{kob1995testing,berthier2010critical} amorphous metal alloys,\cite{zollmer2003diffusion,nowak2017partial} colloidal suspensions,\cite{szamel1991mode,siebenburger2009viscoelasticity,schrack2020mode} polymers,\cite{farago2012mode1,farago2012mode2}  vitrimers,\cite{ciarella2019understanding} and systems interacting through many-body potentials.\cite{ruscher2020glassy}. In all of these systems,  MCT qualitatively captures at least some key features of the glassy dynamics. A natural question is whether GMCT is also applicable for these more complex glass formers. To address this question, we will take the so-called sticky hard sphere model, which is characterized by a hard-core repulsive potential with a short-ranged attractive square well, as our system of study. 

The sticky hard sphere model is the simplest model with both repulsive and attractive interactions but with rich dynamical behavior. In the last 20 years, many simulation\cite{puertas2002comparative,zaccarelli2002confirmation,foffi2002evidence,zaccarelli2003activated,sciortino2003evidence,zaccarelli2004numerical,foffi2004dynamical,saika2004effect,reichman2005comparison,moreno2006anomalous,gonzalez2021mechanical1,gonzalez2021mechanical2} and experimental\cite{mallamace2000kinetic,eckert2002re,pham2002multiple,pham2004glasses,kaufman2006direct,buzzaccaro2007sticky,lu2008gelation,zhang2011cooperative,brown2016correlated} studies have been performed for this model. It is widely accepted that there exists a complex reentrant dynamics when decreasing the temperature of the system at certain fixed values of the  packing fraction and  potential well width. Moreover, in contrast to the typical power-law decays of the intermediate scattering function near the critical liquid-glass point for hard spheres, an anomalous logarithmic decay has been found for sticky hard spheres. Although still debated recently,\cite{fullerton2020glassy} the sticky hard-sphere reentrance is associated with two distinct types of glass: one is the so-called attractive glass, or bonded glass, related to the short deep potential well; the other is the repulsive glass, or non-bonded glass, similar to the hard-sphere glass. In MCT, all the above properties have been successfully predicted. Moreover, MCT predicts a glass-glass transition and even more complex transitions called $A_3$ and $A_4$ singularities, which can also explain the logarithmic decay of the intermediate scattering function. However, we point out that the predictions from MCT are all qualitatively but not quantitatively correct. For example, regarding the phase transition at different well widths, the critical packing fractions are all underestimated by MCT. Furthermore, the $A_3$ and $A_4$ singularity points predicted by MCT have to be rescaled to match the simulation or experimental results. In this work, we revisit the sticky hard sphere model in the context of GMCT, with the aim to test the applicability of GMCT for systems with both attractive and repulsive interactions, and to establish to what extent first-principles predictions for glassy sticky hard spheres can be improved. 

The outline of this paper is as follows. We first introduce the microscopic GMCT framework and the related analytical scaling laws. We also briefly describe the sticky hard sphere model and the numerical details for the application of GMCT to this model. Subsequently, we present the rich phase diagrams of sticky hard spheres predicted from higher-order GMCT and compare them to literature to demonstrate the quantitative improvement attained by GMCT. Next, we discuss the non-ergodicity parameters at the critical points for both the liquid-glass transitions and the glass-glass transition, reflecting the different mechanisms of the two types of glass. The exponent parameters that characterize the dynamics near the corresponding critical points are also provided. Finally, we show the relaxation dynamics near the higher-order $A_3$ and $A_4$ singularities, demonstrating that the pronounced logarithmic decay of the dynamical density correlation functions is preserved within higher-order GMCT. We conclude with some critical remarks and perspectives for future work.

\section{Theory}
\label{sec:theory}
\subsection{Generalized mode-coupling theory}
We first review the microscopic GMCT equations of motion\cite{janssen2015microscopic} and several scaling laws derived from them.\cite{luo2020generalized2}
The microscopic dynamical information of a glass-forming material is assumed to be encoded in the $2n$-point density correlation functions
$F_n(k_1,\hdots,k_n,t)$, defined as 
\begin{equation}
\label{eq:Fndef}
F_n(k_1,\hdots,k_n, t) = \langle \rho_{\bm{-k_1}}(0) \hdots \rho_{-\bm{k_n}}(0)
	\rho_{\bm{k_1}}(t) \hdots \rho_{\bm{k_n}}(t) \rangle,
\end{equation} 
where $\rho_{\bm{k}}(t)=\sum_{j=1}^{N_p} e^{i\bm{k}\cdot\bm{r}_j}/\sqrt{N_p}$ is a collective density mode at wavevector $\bm{k}$ and time $t$, $N_p$ is the total number of particles,
the angle brackets denote an ensemble average, and the label $n$ ($n=1,\hdots,\infty$) specifies the level of the GMCT hierarchy.
Note that when $n=1$, $F_1(k,t)$ is the intermediate scattering function.
In the overdamped limit, the GMCT equations read
\begin{gather} 
	\nu_n\dot{F}_n(k_1,\hdots,k_n,t) + F_n(k_1,\hdots,k_n,t) S^{-1}_n(k_1,\hdots,k_n)J_n(k_1,\hdots,k_n)
	\nonumber \\ 
	+\int_0^t  \dot{F}_n(k_1,\hdots,k_n,t-u) J^{-1}_n(k_1,\hdots,k_n) M_n(k_1,\hdots,k_n,u) du  = 0 \label{eq:GMCTF_n}, 
\end{gather} 
where $\nu_n$ is an effective friction coefficient, 
\begin{equation} 
\label{eq:Sn}
S_n(k_1,\hdots,k_n) \equiv F_n(k_1,\hdots,k_n,t=0) \approx \prod_{i=1}^n S(k_i)
\end{equation}
are $2n$-point static density correlation functions which we approximate as a product of static structure factors $S(k_i)$,
and 
\begin{equation} 
\label{eq:Jn}
J_n(k_1,\hdots,k_n) = \sum_{i=1}^n \frac{D_0k_i^2}{S(k_i)}\prod_{j=1}^n S(k_j)
\end{equation}
 with $D_0$ denoting the bare diffusion constant. 
For the memory functions we have
\begin{eqnarray}
\label{eq:Mn} 
M_n(k_1,\hdots,k_n,t) = \frac{\rho D_0^2}{2} 
\int \frac{d\bm{q}}{(2\pi)^3} \sum_{i=1}^n |V_{\bm{q,k}_i-\bm{q}}|^2
\hphantom{XXXX}
\nonumber \\
\times F_{n+1}(q,|\bm{k}_1-\bm{q}\delta_{i,1}|,\hdots,|\bm{k}_n-\bm{q}\delta_{i,n}|,t), \nonumber \\
\end{eqnarray}
where $\rho$ is the number density, $\delta_{i,j}$ is the Kronecker delta
function, and the $V_{\bm{q,k}_i-\bm{q}}$ are so-called static vertices.
These vertices are defined as
\begin{equation}
\label{eq:V}
V_{\bm{q,k-q}} = 
({\bm{k}} \cdot \bm{q}) c(q) + 
{\bm{k}} \cdot (\bm{k-q}) c(|\bm{k-q}|),
\end{equation}
where $c(q)$ denotes the direct
correlation function.\cite{hansen1990theory} The direct correlation function is related to the static structure factor as $c(q)
\equiv [1-1/S(q)] / \rho$. Therefore, the static structure factor $S(q)$ serves as the
only material-dependent input to the GMCT equations. While the above GMCT equations still rely on several approximations, including factorization of all \textit{static} multi-point correlators [Eq.\ \ref{eq:Sn}] and the neglect of so-called off-diagonal dynamic multi-point correlators,\cite{janssen2015microscopic,szamel2003colloidal,SimonePhdthesis}  we emphasize that the theory is still purely first-principles-based. 

In principle, one should expand the GMCT equations up to $n\rightarrow\infty$ to obtain the correct dynamical density correlations. However, in practice we can only solve the GMCT equations up to a finite level $N<\infty$, which necessitates the use of a closure approximation for the last included level $N$. There are two kinds of GMCT closures that have been well-studied before.\cite{janssen2015microscopic,luo2020generalized1} One kind is the so-called mean-field (MF) closure, which approximates the highest-order correlator $F_N$ as a product of lower-order dynamical density correlation functions. In this paper, we only focus on one such example, for $N>2$,
\begin{eqnarray}
\label{eq:closure_t} 
M_N(k_1,\hdots,k_N,t) =\frac{1}{N-1}
\sum_{i=1}^{N}
M_{N-1}(\{k_j\}^{(N-1)}_{j\neq i},t)F_1(k_i,t) 
\end{eqnarray}
where $\{k_j\}^{(N-1)}_{j\neq i}$ represents the $N-1$ wavenumbers in
$\{k_1,\hdots,k_N\}$ except the $k_i$, and the permutation invariance of all wavenumber arguments $\{k_1,\hdots,k_N\}$ in $M_N(k_1,\hdots,k_N,t)$ has been taken into account. This closure, denoted as MF-$N[(N-1)^11^1]$, is exactly the same as the one used in the original paper deriving the microscopic dynamical GMCT,\cite{janssen2015microscopic} and is qualitatively equivalent to the one used in the paper analyzing the scaling laws for GMCT.\cite{luo2020generalized2} For $N=2$, the closure is trivially satisfied with $F_2(k_1,k_2,t)=F_1(k_1,t)F_1(k_2,t)$, which is in fact the standard-MCT closure \cite{gotze2008complex} that we denote here as MF-$2[1^2]$.
The other kind of GMCT closure, named exponential (EXP-$N$) closure, is a simple truncation of the hierarchy such that $F_N(k_1,\hdots,k_N,t)=0$, which
leads to $F_{N-1}\sim\exp(-t/\tau_{N-1})$ with $\tau_{N-1}=\nu_{N-1}/(S_{N-1}^{-1}J_{N-1})$. It has been demonstrated numerically that the mean-field and exponential closures provide an upper and lower bound for the relaxation dynamics, respectively.\cite{janssen2015microscopic,luo2020generalized1,janssen2016schematic} In the limit $N\rightarrow\infty$, the differences of the relaxation dynamics from these two kinds of closures are expected to disappear. From previous numerical results for hard spheres,\cite{janssen2015microscopic,luo2020generalized1} it is known that mean-field closures require relatively few GMCT levels to predict the emergence of strongly glassy behavior; in this work we therefore only focus on the results with mean-field closures to study the glass transition.

Equations (\ref{eq:GMCTF_n}), (\ref{eq:Mn}), and (\ref{eq:closure_t}) define a unique time-dependent solution $F_n(k_1,\hdots,k_n,t)$ for $n\le N$.\cite{biezemans2020glassy} Before we go to the specific model and the numerical calculation, let us discuss some universal properties of the GMCT solutions which can be obtained from mathematically analyzing the equations.

Firstly, using the Laplace transform and the final value theorem, the long-time limit of the multi-point density correlation function $F_n(k_1,\hdots,k_n)\equiv \lim_{t\to\infty}F_n(k_1,\hdots,k_n,t)$ can be calculated via 
\begin{gather}
F_n(k_1,\hdots,k_n)=S_n(k_1,\hdots,k_n)
\nonumber \\
-\left[S_n^{-1}(k_1,\hdots,k_n)+J_n^{-2}(k_1,\hdots,k_n)M_n(k_1,\hdots,k_n)\right]^{-1},
\label{eq:GMCT_long}
\end{gather}
where $M_n(k_1,\hdots,k_n) \equiv \lim_{t\rightarrow\infty}M_n(k_1,\hdots,k_n,t)$ represents the long-time limit of the memory function. When $n<N$, 
\begin{gather}
M_n(k_1,\hdots,k_n) = \mathcal{M}_n[F_{n+1}]
\nonumber\\
= \frac{\rho D_0^2}{2} 
\int \frac{d\bm{q}}{(2\pi)^3} \sum_{i=1}^n |V_{\bm{q,k}_i-\bm{q}}|^2
 F_{n+1}(q,|\bm{k}_1-\bm{q}\delta_{i,1}|,\hdots,|\bm{k}_n-\bm{q}\delta_{i,n}|), 
 \label{eq:Mn_long} 
\end{gather}
and the MF closure Eq.\ (\ref{eq:closure_t}) becomes 
\begin{gather}
M_N(k_1,\hdots,k_N) =\mathcal{M}_N[M_{N-1},F_{1}]
\nonumber\\
=\frac{1}{N-1}
\sum_{i=1}^{N}
M_{N-1}(\{k_j\}^{(N-1)}_{j\neq i})F_1(k_i),
\label{eq:closure_long} 
\end{gather}
where for notational convenience we have introduced the symbol $\mathcal{M}$ to denote the memory functionals. 
The solutions of Eqs.\ (\ref{eq:GMCT_long}), (\ref{eq:Mn_long}), and (\ref{eq:closure_long}) can exhibit non-trivial singularities that may be analyzed using the Jacobian of these equations.
The Jacobian is equivalent to $\bm{1}-\bm{C}$, where $\bm{C}$ is the stability matrix of Eqs.\ (\ref{eq:GMCT_long}), (\ref{eq:Mn_long}), and (\ref{eq:closure_long}) which satisfies
 $$\bm{H}=\bm{C}\bm{H}.$$
 Here $\bm{H}$ is a vector in the form of $\left[\left\{H_1(k_1)\right\}_{\forall k_1},\left\{H_2(k_1,k_2)\right\}_{\forall k_1,\forall k_2},\hdots,\left\{H_n(k_1,\hdots,k_n)\right\}_{\forall k_1,\hdots,\forall k_n}\right]^T$. Explicitly, for $n<N$, 
\begin{gather}
H_n(k_1,\hdots,k_n)=\left[S_n(k_1,\hdots,k_n)-F_n(k_1,\hdots,k_n)\right]^2J_n^{-2}(k_1,\hdots,k_n)
\nonumber\\
\times\mathcal{M}_n[H_{n+1}].
\end{gather}
and 
\begin{gather}
H_N(k_1,\hdots,k_N)=\left[S_N(k_1,\hdots,k_N)-F_N(k_1,\hdots,k_N)\right]^2J_N^{-2}(k_1,\hdots,k_N)
\nonumber\\
\times
\left\{\mathcal{M}_{N}[M_{N-1},H_1]+\mathcal{M}_N[\mathcal{M}_{N-1}[H_N],F_1]\right\}.
\end{gather}
Since all elements of $\bm{C}$ are non-negative and the positive elements between different levels ensure that the matrix is irreducible, there is a non-degenerate maximum eigenvalue $E$ of the matrix $\bm{C}$ according to the Perron-Frobenius theorem.\cite{meyer2000matrix} The singularities, also called bifurcation points,\cite{arnol2003catastrophe} of the long-time limit of multi-point density correlation functions occur if the Jacobian is a singular matrix, i.e., if the matrix $\bm{C}$ has an eigenvalue $E=1$. In principle, there could be very complex singularities, including a family labeled $A_l$, $l=2,3,...$, which are topologically equivalent to the bifurcation singularities of the real roots of real polynomials of degree $l$.\cite{arnol2003catastrophe} 
The simplest singularity is the so-called $A_2$ bifurcation point, which most reported liquid-glass transitions in the literature belong to. In this class, the 
normalized long-time limits of the dynamic density correlation
functions (also usually called non-ergodicity parameters) $f_n(k_1,\hdots,k_n)\equiv F_n(k_1,\hdots,k_n)/S_n(k_1,\hdots,k_n)$ display a simple bifurcation: For liquid states $f_n(k_1,\hdots,k_n)=0$, while for glass states $f_n(k_1,\hdots,k_n)>0$. Such an $A_2$ singularity is found in e.g.\ the Percus-Yevick hard-sphere system,\cite{luo2020generalized2} for which the only control parameter of the system is the packing fraction $\varphi$; the lowest packing fraction for the glass state, $\varphi^c$, is referred to as the critical point of the glass transition. More generally, if there are $p>1$ control parameters in the system, $f_n(k_1,\hdots,k_n)$ is a function of these $p$ variables and hence in principle there could exist singularities $A_l$ where $l>2$. These are the cases we will pay more attention to in this paper.

At each singularity point, i.e., when $E=1$, the left eigenvector $\hat{e}_n(k_1,\hdots,k_n)$ and right eigenvector $e_n(k_1,\hdots,k_n)$ of the $\bm{C}$ matrix can be obtained with conventions
$$ \sideset{}{'} \sum_{n=1}^{N}  \hat{e}_n(k_1,\hdots,k_n) e_n(k_1,\hdots,k_n) =1 $$
and 
\begin{gather}
\sideset{}{'} \sum_{n=1}^{N} \hat{e}_n(k_1,\hdots,k_n) e^2_n(k_1,\hdots,k_n)\left[S^c_n(k_1,\hdots,k_n)-F^c_n(k_1,\hdots,k_n)\right]^{-1} =1
\end{gather} 
where $\sideset{}{'}\sum_{n=m}$ represents the summation over all possible wavenumbers $k_1,\hdots,k_m$ for level $m$ and the superscript $c$ refers to the critical-point values. The above eigenvectors are used to characterize every transition point by a single number $\lambda$, defined as 
\begin{gather}
\lambda=\sideset{}{'} \sum_{n=N}\hat{e}_N(k_1,\hdots,k_N)\left[S^c_n(k_1,\hdots,k_n)-F^c_n(k_1,\hdots,k_n)\right]^{2}J^{c-2}_n(k_1,\hdots,k_n)
\nonumber\\
\times \mathcal{M}^c_N[\mathcal{M}_{N-1}[e_N],e_1].
\label{eq:lambda}
\end{gather}
For bifurcation singularities of the $A_2$ type, $0<\lambda<1$. This $\lambda$ becomes unity at the end points of the $A_2$ transition curve (or surface), which could be $A_3$ or even higher order singularities.\cite{dawson2000higher,arnol2003catastrophe}

Let us briefly summarize some leading order scaling laws near $A_2$ singularities within the framework of GMCT, both for the normalized long-time limits of the density correlators and for the time-dependent ones [i.e., the solutions of Eqs.\ (\ref{eq:GMCTF_n}), (\ref{eq:Mn}) and (\ref{eq:closure_t})]. Details of these scaling laws can be found in Ref.\  \cite{luo2020generalized2}. 
Here the control parameter is the relative distance to a $A_2$ critical point which, for example, is measured by the packing fraction in hard spheres, $\epsilon=(\varphi-\varphi^c)/\varphi^c$. 
\begin{itemize}
	\item If $\epsilon>0$ and $|\epsilon|\ll 1$, the system is in the glass state and the non-ergodicity parameters scale with $\sqrt{\epsilon}$ as
	\begin{gather}
	f_n(k_1,\hdots,k_n)-f^c_n(k_1,\hdots,k_n)\sim \sqrt{\epsilon} h_n(k_1,\hdots,k_n)
	\label{eq:f_scaling}
	\end{gather} 
	where 
	\begin{gather}
h_n(k_1,\hdots,k_n)=e_n(k_1,\hdots,k_n)/S_n(k_1,\hdots,k_n).
\label{eq:hn}
	\end{gather}
	
	\item If $|\epsilon|\ll 1$, the $\beta$-relaxation regime
	can be characterized by a unique timescale $\tau_\beta\sim |\epsilon| ^ {-1/2a}$. In this regime, the dynamical normalized density correlation functions $f_n(k_1,\hdots,k_n,t)\equiv F_n(k_1,\hdots,k_n,t)/S_n(k_1,\hdots,k_n)$ 
	obey a time-wavenumber factorization property such that 
	\begin{gather}
	f_n(k_1,\hdots,k_n,t)=f^c_n(k_1,\hdots,k_n,t)+h_n(k_1,\hdots,k_n,t)G(t).
	\label{eq:fachG}
	\end{gather}
	Here $G(t)$ is the so-called $\beta$-correlator and it scales with $\epsilon$ as $G(t)\sim \sqrt{\epsilon}g_{\pm}(t/\tau_\beta)$, 
	with the subscript $\pm$ denoting the sign of $\epsilon$.
	For the limiting case that $\epsilon\rightarrow 0^-$, i.e., on the liquid side of the transition, the $\beta$-correlator satisfies
	\begin{gather}
	G(t)\sim\left(\frac{t_0}{t}\right)^{a}\ \ \text{if }t < \tau_{\beta},
	\label{eq:GA}
	\end{gather}
	and
	\begin{gather}
	G(t)\sim -\left(\frac{t}{\tau}\right)^b \ \  \text{if }t > \tau_{\beta},
	\label{eq:GB}
	\end{gather}
	which describes the decay towards and away from the plateau, 
	respectively. The latter equation for the late $\beta$-relaxation regime is also called the von Schweidler law. The exponents $a$ and $b$ can be determined by $\lambda$ in Eq.\ (\ref{eq:lambda}) via 
	\begin{equation}
	\lambda=\Gamma(1-a)^2/\Gamma(1-2a)=\Gamma(1+b)^2/\Gamma(1+2b).
	\label{eq:ablambda}
	\end{equation}
	The $t_0$ in Eq.\ (\ref{eq:GA}) is a time scale characterizing transient dynamics,  while $\tau$ in Eq.\ (\ref{eq:GB}) is the $\alpha$-relaxation time. Within GMCT under MF closures, we have $\tau\sim |\epsilon|^{-\gamma}$ with 
	\begin{equation}
	\gamma=1/2a+1/2b.
	\label{eq:gammaab}
	\end{equation}
	
	\item If $\epsilon<0$ and $|\epsilon|\ll 1$, the system is in the liquid state and the dynamical normalized density correlation functions in the $\alpha$-relaxation regime obey a
	time-density superposition principle 
	\begin{gather}
	f_n(k_1,\hdots,k_n,t)=\tilde{f_n}(k_1,\hdots,k_n,t/\tau).
	\label{eq:alphascaling}
	\end{gather}
\end{itemize}
All the above scaling laws near $A_2$ singularities are very similar to those predicted from  standard MCT, i.e., GMCT with MF closure level $N=2$. However,  we point out that in higher-order GMCT i) all the exponents ($a$, $b$, $\lambda$, $\gamma$, etc.) become closure-level dependent, and ii) the dynamics of multi-point density correlators can also be predicted. In particular, the latter might contain spatiotemporal information related to dynamical heterogeneity. 

For higher-order singularities $A_l$ with $l>2$, a major difference with the scaling laws above is that near such singularities, to leading order in $\sqrt{|\epsilon|}$, the $\beta$-correlators $G(t)$ may follow a logarithmic decay \cite{gotze1989logarithmic,dawson2000higher} 
, i.e. 
\begin{gather}
G(t)\sim -\sqrt{|\epsilon|}\ln (t/\tau_2).
\label{eq:Gtlog}
\end{gather} 
The technique to obtain this scaling law is similar to the one for the $g_\pm(t/\tau)$ near  $A_2$ singularities (see Ref.\ \cite{luo2020generalized2}
for more details.) The difference is that when $\lambda=1$, the equation of the $\beta$-correlator becomes 
\begin{gather}
\sigma|\epsilon|-s^2G^2(s)+s\mathcal{L}\left[G^2(t)\right]=0
\end{gather}
where $\sigma$ is a constant that depends on the system and on the GMCT closure. If $\sigma<0$, the solution is Eq.~(\ref{eq:Gtlog}), similar to  standard MCT.\cite{gotze2002logarithmic}
Therefore, the emergence of logarithmic decay can be used as an indicator for the existence of high-order singularities, both in simulation and experiment.\cite{sciortino2003evidence} Another difference near higher-order singularities is that the superposition principle in the $\alpha$-relaxation regime need not be applicable anymore, which is related to the breakdown of the von Schweidler law in the late $\beta$-relaxation regime.
In Sec.\ \ref{sec:res}, we will discuss more subtle effects for control parameters $p=3$ where $A_l$ ($l=2,3,4$) singularities exist.

\subsection{Sticky hard sphere model}
\begin{figure}
\centering
  \includegraphics[width=8.3cm]{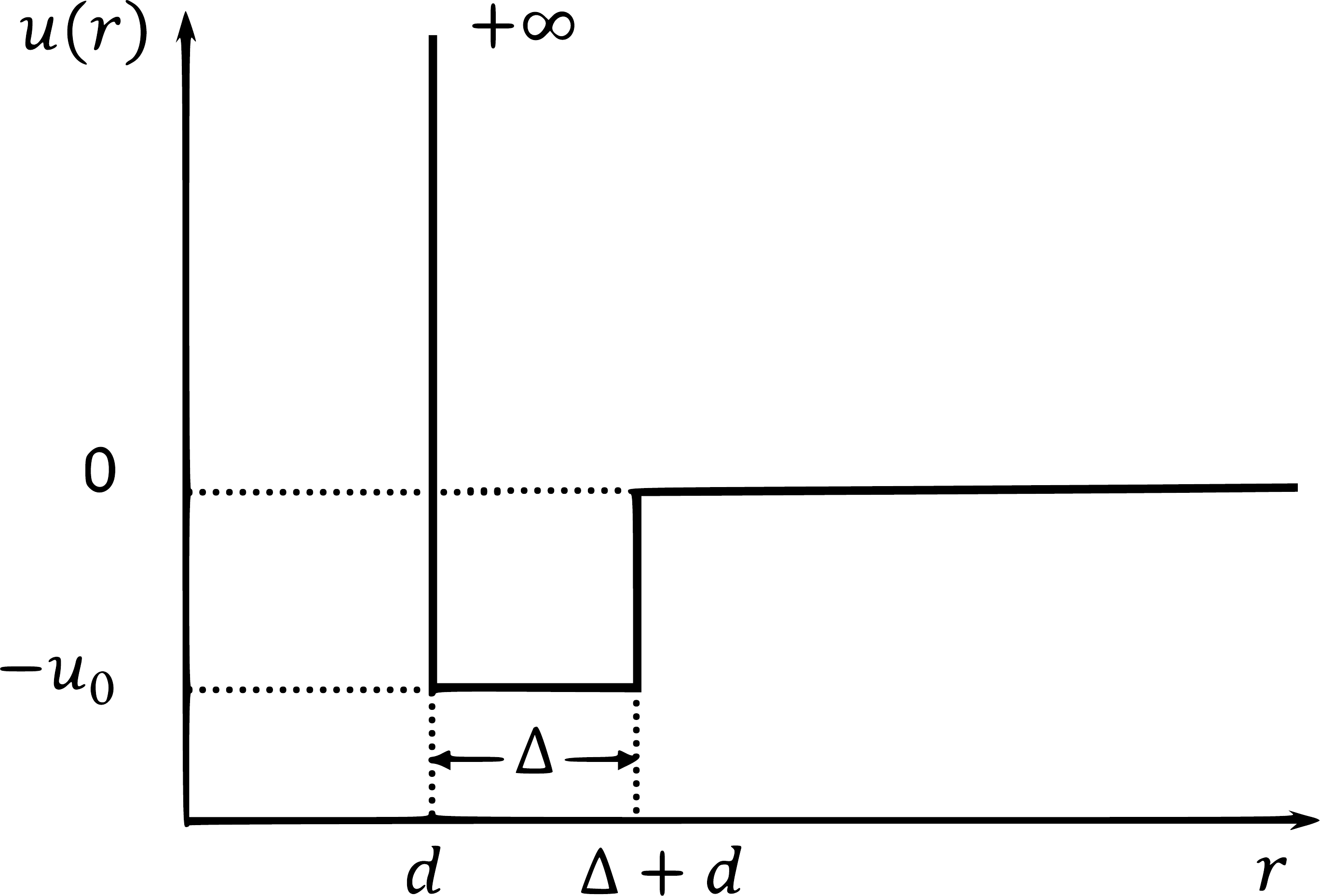}
  \caption{Pair potential $u(r)$ for sticky hard spheres as a function of interparticle distance $r$.}
  \label{fig:shs_potential}
\end{figure}
We apply our GMCT framework to a system composed of sticky hard spheres. The pair interaction potential of this system includes both a repulsive and attractive part, as shown in Fig.~\ref{fig:shs_potential}. Within the particle diameter $d$, the value of the potential is $u(r)=+ \infty$, which accounts for hard-core repulsion to prevent particle overlap. Outside the hard-core region there is a small square potential well with width $\Delta \ll d$, i.e., for $d<r<d+\Delta $ we have $u(r)=-u_0$, with $u_0>0$. Thus, if the distance of two particles falls within this range, the particles will attract each other to be sticky. When $r>d+\Delta$, $u(r)=0$.

There are three independent parameters of this model. One is the packing fraction $\varphi=\pi \rho d^3/6$. The second is the reduced temperature $\theta$, defined as the inverse of the attraction well depth over thermal energy, i.e., $\theta=k_BT/u_0$. The third one is the relative width of the well, $\delta=\Delta/d$. Note that when $u_0=0$ ($\theta\rightarrow\infty$), this sticky hard sphere model reduces to the hard sphere model and the only parameter is the packing fraction $\varphi$. We calculate the structure factors $S(k)$ from the pair potential using the Ornstein-Zernike equation together with the mean-spherical approximation.\cite{hansen1990theory} More specifically, we use the approach motivated by Baxter's theory in Ref.\ \cite{dawson2000higher} 
to numerically obtain the structure factors in the three-dimension control parameter space $(\varphi, \theta, \delta)$; these structure factors subsequently serve as our GMCT input. 
 
 For the GMCT calculations, we numerically solve Eqs.\ (\ref{eq:GMCTF_n})--(\ref{eq:closure_t}) for the time-dependent dynamics of the multi-point density correlation functions, and Eqs.~(\ref{eq:GMCT_long})--(\ref{eq:closure_long}) for the corresponding long-time limits, for MF closure levels $N=2,3,4$. 
 The wavevector-dependent integrals over $\bm{q}$ in the memory functions are approximated
 as a double Riemann sum\cite{franosch1997asymptotic} with
 an equidistant wavenumber grid of $N_k=500$ points that ranges from $kd=0.2$ to 
 $kd=200.2$. Such a fine wavenumber grid is necessary to reach convergence, in particular for low values of $\theta$.
 For the time-dependent solutions
 we use the integration algorithm described by Fuchs \textit{et al.},\cite{fuchs1991comments} 
 starting with a time step size of $\Delta t=10^{-6}$ that is subsequently doubled every 32 points. 
 In all our GMCT calculations we set the bare diffusion constant $D_0=1$ and the effective friction coefficients $\nu_n=1$ for $1\le n\le N$.
 
\section{Results and discussion}
\label{sec:res}
\subsection{Phase diagram}
\label{sec:phasediagram}
\begin{figure}[h!]
	\centering
	\includegraphics[width=8.3cm]{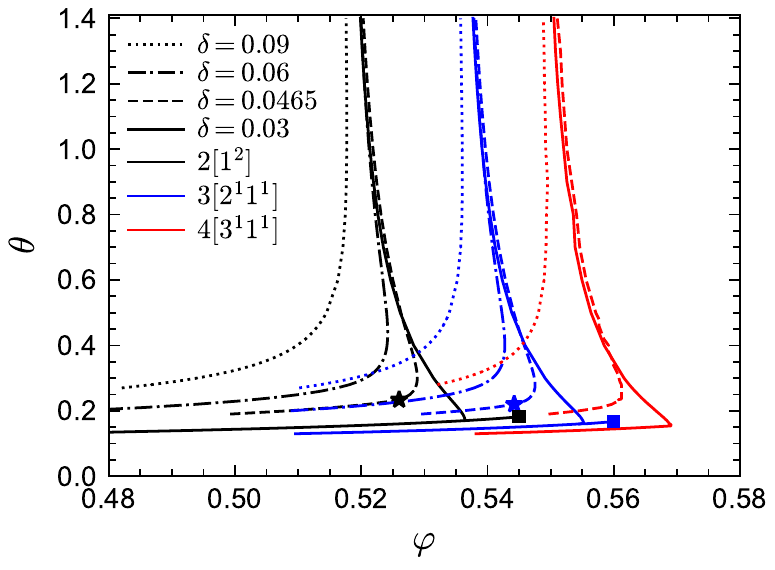}
	\caption{Phase diagrams of sticky hard spheres, obtained from GMCT with MF closure levels $N=2,3,4$. The glass transition curves are shown as a function of packing fraction $\varphi$ and reduced temperature $\theta$ for several values of the relative attraction well width $\delta$. The $A_3$ end points are marked by squares and the $A_4$ end points are marked by stars. The $A_3$ and $A_4$ points also exist for GMCT closure level $N=4$, but their positions could not be accurately determined due to limiting computing power.}
	\label{fig:phase_diagram}
\end{figure}

We first report the GMCT phase diagrams in the $(\varphi,\theta)$ plane for several fixed values of $\delta$ (i.e., fixed potential well widths), as shown in Fig.\ \ref{fig:phase_diagram}. Let us start by discussing the $N=2$ results, which are also fully consistent with the standard-MCT results of approximately 20 years ago.\cite{dawson2000higher} For all values of $\delta$ considered ($\delta=0.09, 0.06, 0.0465, 0.03$), a common phenomenon is that for a given reduced temperature $\theta$, there is always a liquid-glass transition point $\varphi^c$. That is, for $\varphi>\varphi^c$ the non-ergodicity parameters are larger than zero and the system is in the glass state, while for $\varphi<\varphi^c$ the non-ergodicity parameters are all $0$, representing a liquid state. Thus, for an arbitrary temperature and potential well width, one may always find an ergodicity-breaking transition by going to  sufficiently high packing fractions. 
However, for a given packing fraction, the influence of the reduced temperature is more subtle and depends on the width of the potential well. At relatively large widths, such as $\delta=0.09$, the picture is fairly trivial: the lower the temperature, the more glassy the system. This is indeed similar to systems with a relatively large attractive potential well, such as the Lennard-Jones potential.\cite{kob1995testing} Conversely, when the width of the attractive well decreases to e.g.\  $\delta=0.06$, the scenario changes: at high temperatures the system is in the glass state, at lower temperatures the glass melts into a liquid state, and finally at even lower temperatures the system re-enters the glass phase. This is the famous reentrant phenomenon of systems with a short attractive well, and can be found regardless of the shape of the well.\cite{gotze2003higher} Physically, the existence of the glass phase at high reduced temperatures (i.e., the repulsive glass) can be understood in terms of the cage effect, similar to the case of hard spheres. That is, the MCT modes with wavelengths close to the average distance of neighbors play the dominant role. By contrast, the glass phase at low temperatures (i.e., the so-called attractive glass) arises from particle bonding, with the MCT modes at wavelengths comparable to the width of the attractive potential well becoming more important. The liquid state in between can be explained by the following two arguments:\cite{pham2002multiple,dawson2000higher} i) compared to the repulsive glass, the attraction causes the separation of two particles to be smaller so that the average size of the free space increases, facilitating  longer-distance particle motion and thus faster structural relaxation; ii) compared to the attractive glass, the "stickiness" between particles is weaker, hence the particles have a higher probability to break their bonds and move relatively fast away from their initial positions.

The two-glass picture becomes a glass-glass transition for very small potential well widths. For example, when $\delta=0.03$, in addition to the reentrant liquid-glass transition, there is another $A_2$ glass-glass transition predicted from MCT (black solid line from the crossing point near $\varphi=0.535$ to the black square marker in Fig.~\ref{fig:phase_diagram}). The existence of this glass-glass transition is still debated, most recently based on swap Monte Carlo simulations,\cite{fullerton2020glassy} but from the theoretical point of view, this transition indeed exists since the maximum eigenvalue $E$ of the $\bm{C}$ matrix introduced in the previous section is unity at the glass-glass transition curve, with $\lambda<1$. Moreover, when $\delta=0.03$, along the $A_2$ glass-glass transition curve, $\lambda$ increases with increasing $\varphi$ until $\lambda\rightarrow 1$ (black solid line in Fig.~\ref{fig:lambda}) indicating an end point with an $A_3$ singularity (black square in Fig.~\ref{fig:phase_diagram}). When increasing the width $\delta$, the length of the glass-glass transition curve shrinks and there will be a $\delta^*$ at some $\varphi^*$ and $\theta^*$ when the curve vanishes. The point $(\varphi^*,\theta^*,\delta^*)$ is the $A_4$ singularity (black star in Fig.~\ref{fig:phase_diagram}).
 
 While standard MCT already successfully captures the two competing effects, i.e., caging and bonding, underlying vitrification in sticky hard spheres, the MCT-predicted phase diagram is not quantitative accurate. Indeed, as in many other systems, MCT generally tends to overestimate the critical packing fraction  of the liquid-glass transition point. Let us therefore turn to the higher-order GMCT results in Fig.~\ref{fig:phase_diagram}. 
 We find that overall the shape of the phase diagram under higher-order MF closures remains similar to the MCT case, with all the liquid-glass transitions, glass-glass transitions and $A_l$ ($l=2,3,4$) singularities preserved. However, when the MF closure level $N$ in GMCT increases, all the phase transition curves for a given $\delta$ move to larger packing fraction values. Note also that at high reduced temperatures $\theta$, the critical liquid-glass transition points at different widths $\delta$ will converge to the corresponding GMCT critical points for hard spheres. It has been shown that the GMCT-predicted critical packing fraction $\varphi^c$ for hard spheres shifts toward higher values in a seemingly convergent manner, i.e., the difference $\varphi^c(N)-\varphi^c(N-1)$ becomes increasingly smaller as $N$ increases.\cite{luo2020generalized1} The phase transition curves in Fig.~\ref{fig:phase_diagram} also imply this uniform convergence for any given $\delta$ and $\theta$.  This is arguably one of the biggest triumphs of GMCT: by systematically including higher levels of density correlators in the GMCT hierarchy, the critical point can be systematically improved, even in the presence of competing repulsive and attractive particle interactions. 

\begin{figure}[h!]
	\centering
	\includegraphics[width=8.3cm]{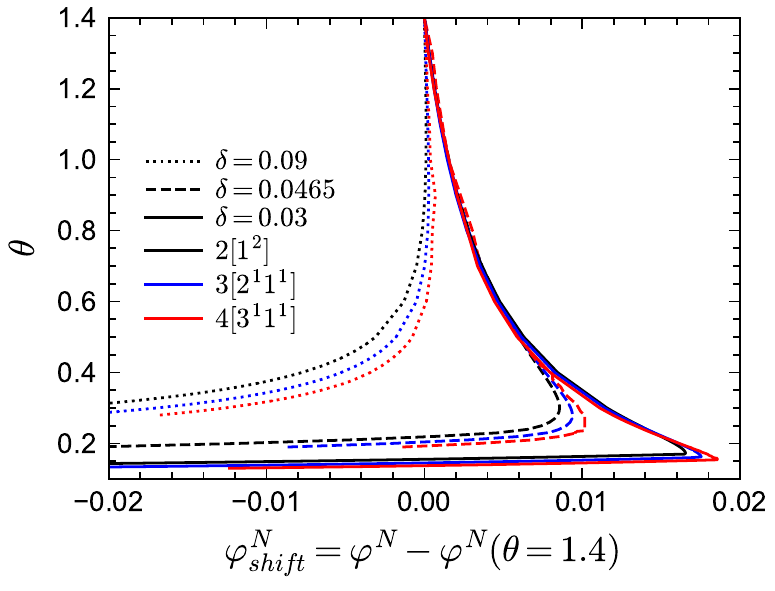}
	\caption{Shifted liquid-glass transition lines of sticky hard spheres for different values of $\delta$ and different GMCT MF closure levels. The reference critical packing fraction for each closure level $N$ is chosen at the reduced temperature of $\theta=1.4$, which is high enough to be regarded as the liquid-glass transition point for hard spheres.}
	\label{fig:phi_shift_all}
\end{figure}

\begin{figure}[h!]
	\centering
	\includegraphics[width=8.3cm]{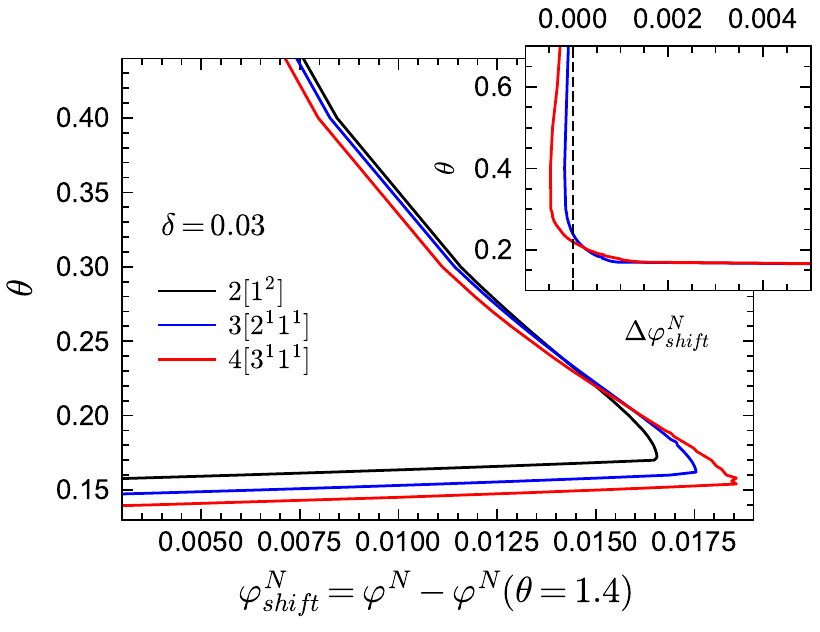}
	\caption{Shifted liquid-glass transition lines of sticky hard spheres for different GMCT MF closure levels at a relative attraction well width of $\delta=0.03$. The results are the same as in Fig.~\ref{fig:phi_shift_all} but enlarged for the low reduced-temperature region. The inset shows the relative difference between higher-order GMCT and MCT, i.e.\  $\Delta\varphi^{N=3}_{shift}\equiv \varphi^{N=3}_{shift}-\varphi^{N=2}_{shift}$ (blue curve)
	and $\Delta\varphi^{N=4}_{shift}\equiv \varphi^{N=4}_{shift}-\varphi^{N=2}_{shift}$ (red curve).
	}
	\label{fig:phi_shift}
\end{figure}

In practice, the phase transition lines from MCT are usually shifted or rescaled to compare with simulation or experimental results. In Fig.~\ref{fig:phi_shift_all} we present the shifted GMCT liquid-glass transition curves for $N=2,3,4$ relative to the corresponding critical points $\varphi^N(\theta=1.4)$ at each $\delta$. We have chosen $\theta=1.4$, which is approximately equivalent to the hard-sphere limit, as a reference point for convenience. For each $\delta$ value, it is obvious that the shifted phase transition curves at different closure levels $N$ do not collapse to a single curve, especially at low temperatures $\theta$. This indicates that the GMCT-predicted phase diagrams under higher closure levels are non-trivial in the sense that they cannot be obtained by simply shifting the MCT results. At high $\delta$, such as $\delta=0.09$, increasing $N$ leads to larger shifted critical packing fractions $\varphi^N_{shift}$ of the liquid-glass transition (dotted lines in Fig.~\ref{fig:phi_shift_all}). Moreover, the difference between the $\varphi^N_{shift}$ with $N=3$ (or $N=4$) and the one with $N=2$, i.e., $\Delta\varphi^{N=3}_{shift}\equiv \varphi^{N=3}_{shift}-\varphi^{N=2}_{shift}$, is larger at low $\theta$. This implies that the higher order density correlations included in GMCT may contribute more to the dynamics when the attractive potential well is deeper. However, when $\delta$ decreases to e.g.\ $\delta=0.03$ (solid lines in Fig.~\ref{fig:phi_shift_all} and Fig.~\ref{fig:phi_shift}), the relation between the temperature $\theta$ and the shifted liquid-glass transition lines becomes more complex. At low temperatures, when $N$ increases, the  shifted critical packing fractions $\varphi^N_{shift}$ increase as well, which is similar to the $\delta=0.09$ case. However, at higher temperatures,  $\varphi^N_{shift}$ decreases as the level $N$ increases. There thus exists a crossing point between any two shifted phase transition curves with different closure levels $N$; this crossing point $(\theta\approx 0.22, \varphi^N_{shift}\approx 0.015)$ is close to the 'convex' part of the phase diagram, and more specifically at the high-temperature side of the convex region. 
The $\Delta\varphi^{N}_{shift}$ shown in the inset of Fig.~\ref{fig:phi_shift} demonstrates this opposite trend for relatively high and low temperatures.

This subtle, complex effect at small $\delta$, which in our theory emerges only when going beyond the standard-MCT level ($N>2$), is in fact also seen in earlier empirical studies of sticky hard spheres dating back to at least 20 years ago, but it appears to have been ignored thus far, or was perhaps regarded as a small (numerical) error. For example, in Fig.~1 of the simulation work by Sciortino, Tartaglia, and Zaccarelli,\cite{sciortino2003evidence}, the simulation data at high $k_BT/u_0$ have a smaller packing fraction than the shifted MCT predicted liquid-glass transition curve, while the data points at low $k_BT/u_0$ are at higher packing fraction values compared to the predicted curve. This trend fully agrees with our predictions from GMCT with higher closure levels $N$. Another example is from the experimental work by Pham \textit{et al.}\cite{pham2002multiple} Although the experimental interaction potential is not the same as the one shown in Fig.~\ref{fig:shs_potential}, both contain a short-ranged attractive well, so that we may expect the effect of higher-order GMCT compared to standard MCT to be similar. Indeed, at high polymer concentrations, which corresponds to  low reduced temperatures in our present work, the rescaled phase transition curve in Fig.~1 of Ref.\ \cite{pham2002multiple} 
has lower packing fractions than the experimental data, while at low polymer concentrations, the critical packing fractions are overestimated by the MCT predictions after rescaling. Notice that the crossing point of the shifted MCT predicted curve and the phase transition curve drawn from data in both simulation and experiment is at the side of high reduced temperatures (high $k_BT/u_0$ in the simulation and low polymer concentration $c_p$ in the experiment), which further demonstrates that the improvement of the liquid-glass transition phase diagram made by higher-order GMCT is reasonable.

\subsection{Non-ergodicity parameters}

\begin{figure*}[h!]
	\centering
	\includegraphics[width=17.1cm]{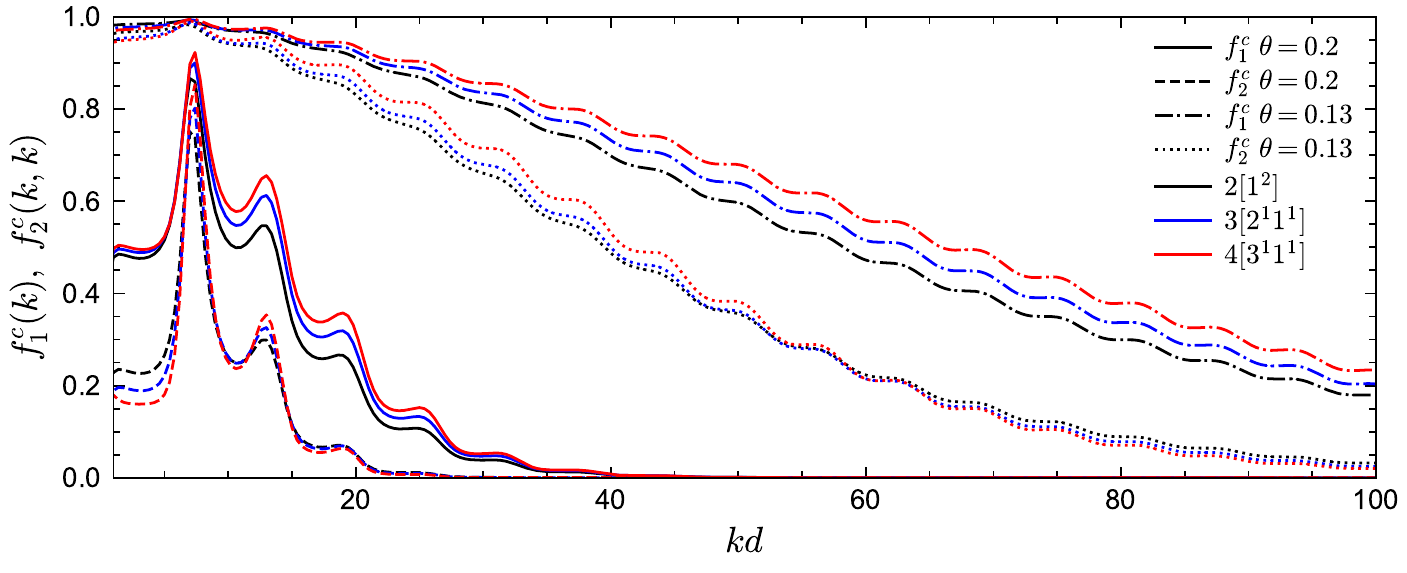}
	\caption{Non-ergodicity parameters $f^c_1(k)$ and $f^c_2(k,k)$ as a function of wavenumber $k$ at the liquid-glass transition point for different GMCT MF closure levels at $\delta=0.03$. At $\theta=0.2$, the critical packing fractions are $\varphi^c=0.53571,\ 0.55382,\ 0.56667$ for closure levels $N=2,\ 3,\ 4$, respectively; at $\theta=0.13$, $\varphi^c=0.46714,\ 0.50959,\ 0.53817$ for $N=2,\ 3,\ 4$.}
	\label{fig:fc}
\end{figure*}

We now turn to the long-time limits of the normalized multi-point density correlation functions, i.e., the non-ergodicity parameters $f_n(k_1,\hdots,k_n)$. We will focus on a very low attractive potential well width of $\delta=0.03$, for which the sticky hard sphere model exhibits both glass-liquid-glass reentrance and a glass-glass transition. 
Let us first consider the liquid-glass transition. Figure \ref{fig:fc} shows the critical non-ergodicity parameters $f^c_1(k)$ and $f^c_2(k,k)$ at two different temperatures, one at the critical repulsive glass state ($\theta=0.2$) and the other at the critical attractive glass state ($\theta=0.13$). 
It can be seen that the repulsive glass state is similar to the hard-sphere case in the sense that $f^c_1(k)$ is strongly modulated by the shape of $S(k)$ and the localization length is around the size of one particle diameter.\cite{fabbian1999ideal,bergenholtz1999nonergodicity,dawson2000higher,luo2020generalized1} Conversely, the non-ergodicity parameters of the attractive glass are much higher than those of the repulsive glass, and they decay to zero over a much wider range of wavenumbers $k$.\cite{fabbian1999ideal,bergenholtz1999nonergodicity,dawson2000higher} This indicates that particle bonding at small distances indeed plays a more important role in the attractive glass phase. Notably, we observe this phenomenology for all considered GMCT closure levels $N$. Hence, even though higher-order GMCT  predicts quantitatively different values of the non-ergodicity parameters, the theory clearly distinguishes between the two  types of glasses in a similar manner as standard MCT. 


Let us further quantify the difference between MCT and higher-order GMCT by careful inspection of Fig.\ \ref{fig:fc}. For the lowest-order non-ergodicity parameters $f_1^c(k)$, we find that increasing $N$ generally leads to an increase in $f_1^c(k)$, similar to what has been reported previously for GMCT of hard spheres (see Fig.~1 of Ref.\ \cite{luo2020generalized2}). The only small exception to this $N$-dependent trend for sticky hard spheres is observed at low temperature ($\theta=0.13$) and at small wavenumbers $kd<7.4$, where $k_0d=7.4$ corresponds to the first peak of $S(k)$. Overall, at the level of the two-point density correlators, the  effect of higher-order GMCT corrections to MCT thus appears to be qualitatively the same for both the attractive and repulsive glass.  However, the influence of $N$ on the diagonal four-point density correlators $f^c_2(k,k)$ is more complex. Indeed, for large wavenumbers $k$ ($kd>15$ for $\theta=0.2$ and $kd>60$ for $\theta=0.13$), $f^c_2(k,k)$ monotonically decreases with increasing $N$ for both glass types, while for smaller $k$, the ordering of the $f^c_2(k,k)$ curves changes under different closure levels. Moreover,  the number of such crossovers is different for different temperatures. Hence, in general we can conclude that within GMCT $f^c_2(k,k)\neq f_1^c(k)\times f_1^c(k)$, a result that may hint at the presence of dynamical heterogeneities.


\begin{figure*}[h!]
	\centering
	\includegraphics[width=17.1cm]{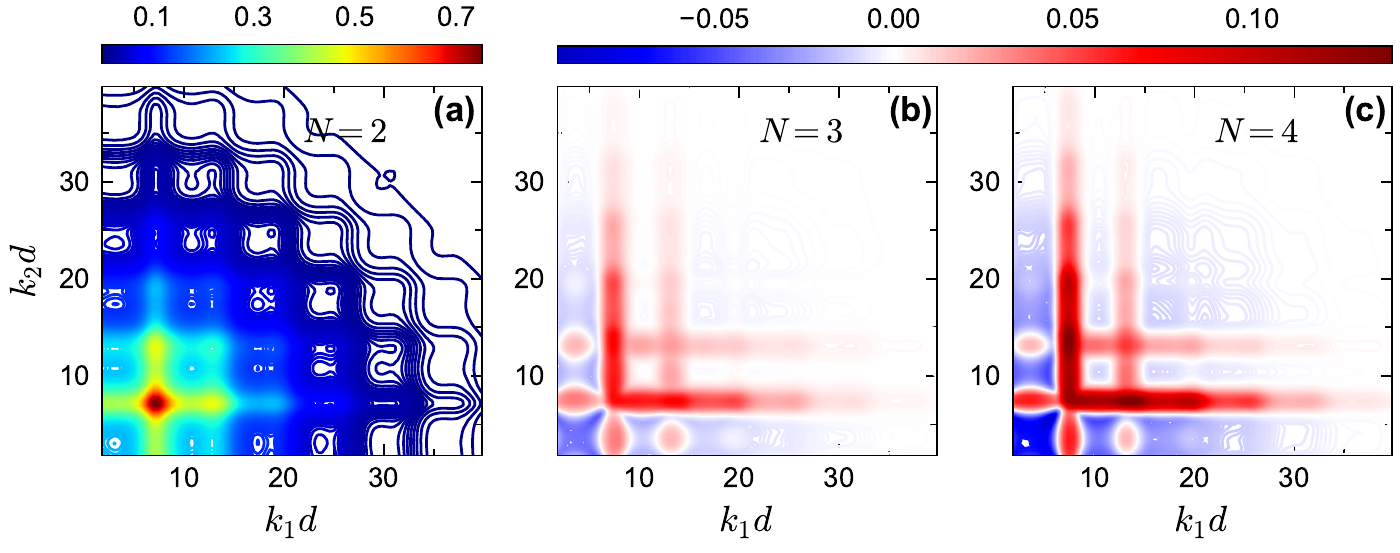}
	\caption{(a) Non-ergodicity parameters $f_2^c(k_1,k_2)$ as a function of wavenumbers $k_1$ and $k_2$ at the critical packing fraction for closure level $N=2$ (i.e., standard MCT) at $\theta=0.2$ and $\delta=0.03$. The patterns of $f^c_2(k_1,k_2)$ from $N=3$ and $N=4$ are similar. (b) The difference of the non-ergodicity parameters $f^c_2(k_1,k_2)$ of closure level $N=3$ with respect to $N=2$. (c) The difference of the non-ergodicity parameters $f^c_2(k_1,k_2)$ of closure level $N=4$ with respect to $N=2$. The packing fractions are the same as in Fig.\ \ref{fig:fc}.}
	\label{fig:fc2d_theta02}
\end{figure*}

\begin{figure*}[h!]
	\centering
	\includegraphics[width=17.1cm]{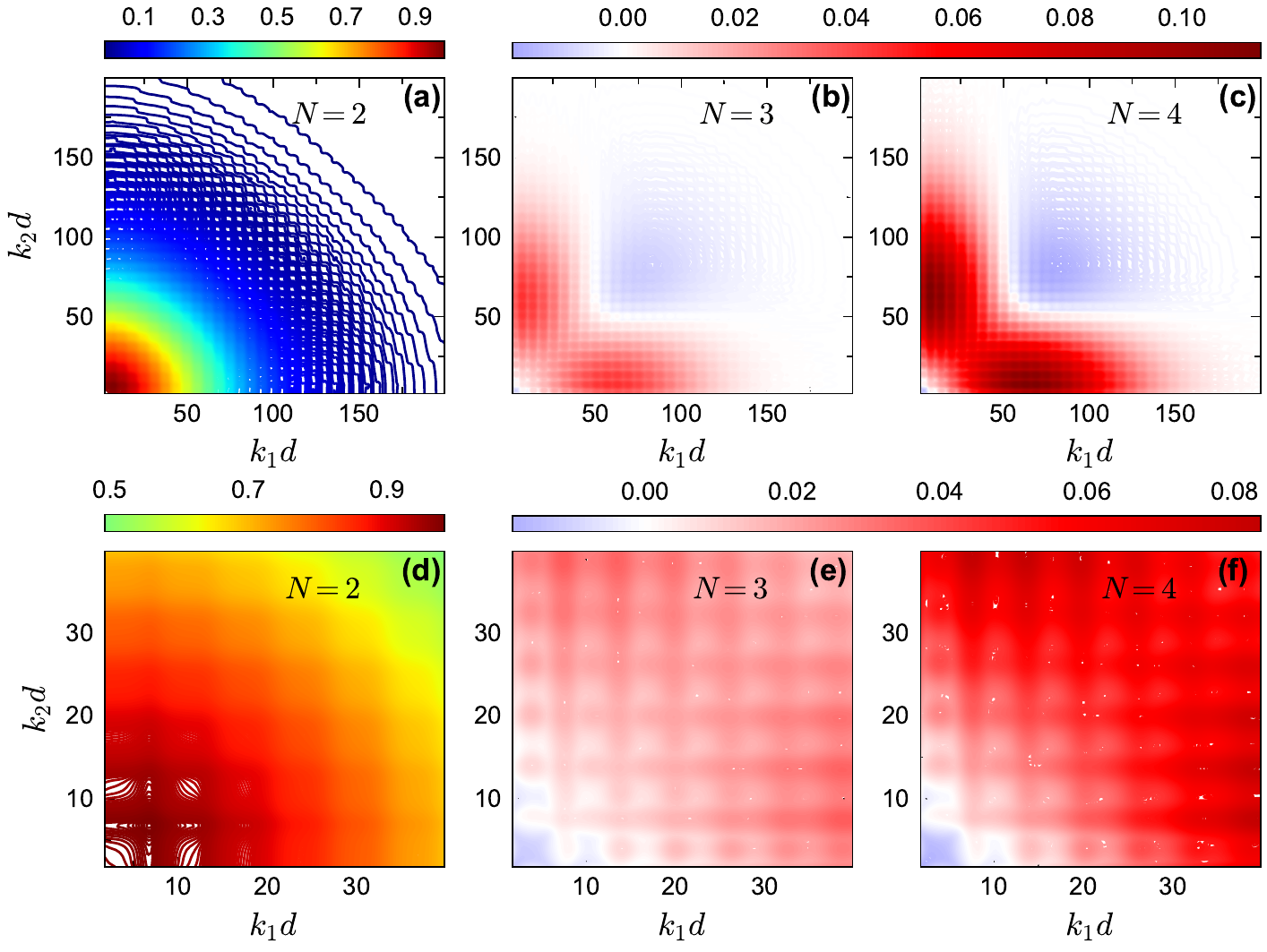}
	\caption{(a) Non-ergodicity parameters $f_2^c(k_1,k_2)$ as a function of wavenumbers $k_1$ and $k_2$ at the critical packing fraction for closure level $N=2$ (i.e., standard MCT) at $\theta=0.13$ and $\delta=0.03$. The patterns of $f^c_2(k_1,k_2)$ from $N=3$ and $N=4$ are similar. (b) The difference of the non-ergodicity parameters $f^c_2(k_1,k_2)$ of closure level $N=3$ with respect to $N=2$. (c) The difference of the non-ergodicity parameters $f^c_2(k_1,k_2)$ of closure level $N=4$ with respect to $N=2$. Panels (d)--(f) represent the zoomed in small-wavenumber regime of panels (a)--(c), respectively.  The packing fractions are the same as in Fig.\ \ref{fig:fc}.}
	\label{fig:fc2d_theta013}
\end{figure*}

 
The off-diagonal correlators $f^c_2(k_1,k_2)$ at $\theta=0.2$ and $\theta=0.13$ are shown in Fig.~\ref{fig:fc2d_theta02}(a) and Fig.~\ref{fig:fc2d_theta013}(a)(d), respectively. We find that the pattern of $f^c_2(k_1,k_2)$ is qualitatively rather similar for all closure levels $N$, with a shape that is modulated by $S(k)$. For MCT this shape is in fact trivial since then $f^c_2(k_1,k_2)=f^c_1(k_1)\times f^c_1(k_2)$; for higher-order GMCT, our repulsive-glass results are also consistent with the GMCT results for hard spheres.\cite{janssen2015microscopic} To see the effect of $N$ for sticky hard spheres more clearly, we plot the difference between $f^c_2(k_1,k_2)$ for $N>2$ and $N=2$ in Fig.~\ref{fig:fc2d_theta02}(b)(c) and Fig.~\ref{fig:fc2d_theta013}(b)(c)(e)(f).  
It is clear that as $N$ increases, the GMCT deviations of $f^c_2(k_1,k_2)$ with respect to MCT increase both in absolute value and in wavenumber range. 
However, comparing the results for the repulsive and attractive glass, we see that for the attractive glass phase the deviations extend over a much broader window of wavenumbers. We speculate that this difference may stem from the different physical mechanisms, i.e., caging versus bonding, underlying the vitrification process, and it might also point toward different degrees of dynamical heterogeneity in the two glassy phases. This hypothesis, however, still requires further study and should be investigated in future work. 

\begin{figure}[h!]
	\centering
	\includegraphics[width=8.3cm]{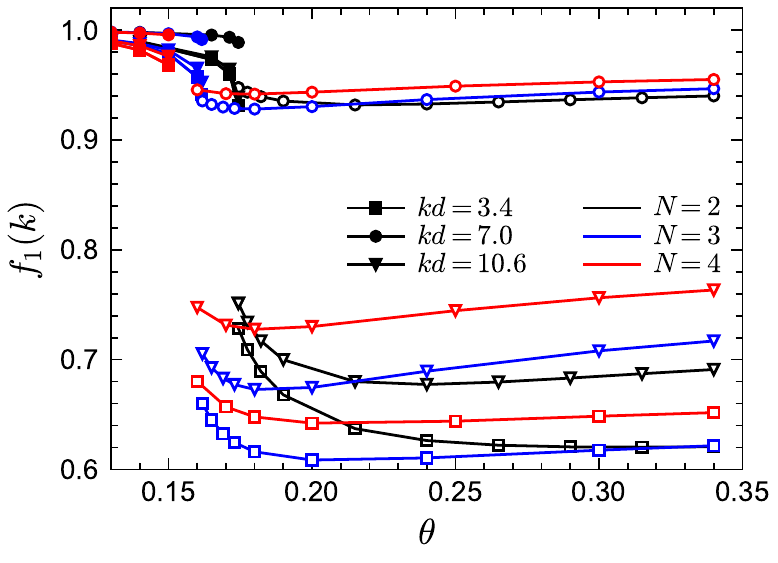}
	\caption{Non-ergodicity parameters $f_1(k)$ as a function of the reduced temperature $\theta$ for fixed $\delta=0.03$ and fixed packing fraction $\varphi$.  The packing fraction values are selected in such a way that the path along $\theta$ crosses the glass-glass transition curve for each corresponding GMCT closure level $N$. For $N=2$, the packing fraction is $\varphi=0.54$; for $N=3$, $\varphi=0.556$; for $N=4$, $\varphi=0.57$. Filled and open symbols represent the attractive and repulsive glass, respectively.}
	\label{fig:fc_vs_theta}
\end{figure}

To demonstrate the existence of the glass-glass transition within GMCT, we plot the non-ergodicity parameters $f_1(k)$ as a function of $\theta$ for different closure levels $N$. Figure \ref{fig:fc_vs_theta} shows $f_1(k)$ for three selected packing fractions and for three representative wavenumbers $k$; we mention that the wavenumber dependence of $f_1(k)$ at both sides of the glass-glass transition line is similar to the result of Fig.~\ref{fig:fc}.
For all closure levels $N$ and all considered wavenumbers $k$, we can see that  $f_1(k)$ initially decreases as $\theta$ increases, until a discontinuous drop occurs; this point corresponds to the $A_2$ singularity of the attractive-to-repulsive-glass transition. In this process, $f_1(k)$ follows the same scaling law as Eq.\ (\ref{eq:f_scaling}). As $\theta$ further increases within the repulsive glass phase, $f_1(k)$ will decrease further for all $N$ (see e.g.\ the open triangles in Fig.~\ref{fig:fc_vs_theta}). However,  after crossing a certain temperature, $f_1(k)$ is seen to increase with $\theta$ and finally approaches the hard-sphere limit. This effect is essentially the counterpart of the glass-liquid-glass reentrance within the glassy region. The non-trivial variation of $f_1(k)$ with $\theta$ near but above the critical temperature is a precursor phenomenon of the nearby $A_3$ singularity. Therefore, we conclude that for all GMCT mean-field closure levels $N$, the glass-glass transition and the $A_3$ singularity are preserved but with quantitatively changed positions in the three-dimension parameter space of sticky hard spheres.

\begin{figure}[h!]
	\centering
	\includegraphics[width=8.3cm]{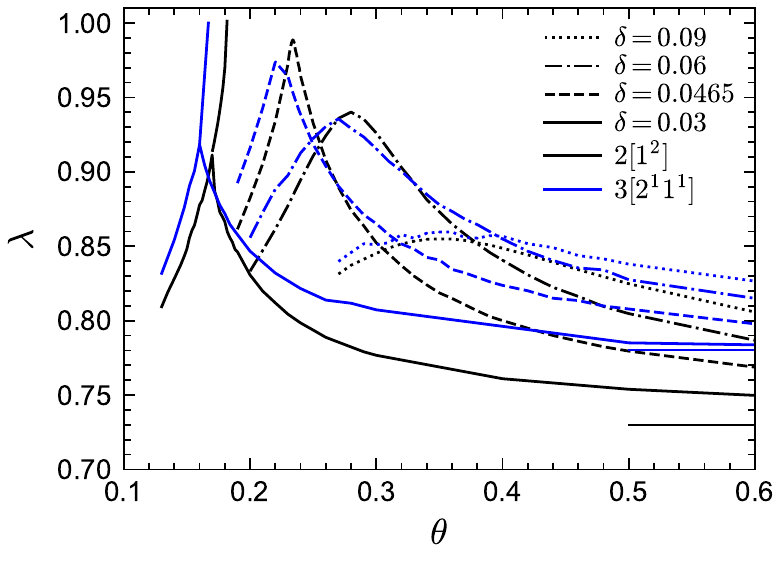}
	\caption{Exponent parameter $\lambda$ for points on the transition lines for GMCT MF closure levels $N=2$ and $N=3$. The thin horizontal black and blue lines indicate the asymptotic hard-sphere values for $N=2$ ($\lambda=0.73$) and $N=3$ ($\lambda=0.78$), respectively.}
	\label{fig:lambda}
\end{figure}

Finally we discuss the exponent parameter $\lambda$, which plays a crucial role in the context of $A_l$ singularities. Figure \ref{fig:lambda} shows  $\lambda$ as a function of $\theta$ for the phase transition lines under MF closure levels $N=2$ and $N=3$. Let us first focus on the standard MCT ($N=2$) results. For relative high values of $\delta$, such as $\delta=0.09,0.06$, $\lambda$ shows a single maximum. As $\delta$ decreases, the peak value of $\lambda$ increases and the corresponding position of $\theta$ decreases. However, for very small $\delta$, e.g.\ $\delta=0.03$, there is another line, representing the glass-glass transition, above the peak. This line ends at the $A_3$ singularity with $\lambda=1$. Therefore there must be a $\delta$ between $0.06$ and $0.03$, which we estimate to be $\delta\approx0.0465$, for which the glass-glass transition curve shrinks to zero and the peak lies at $\lambda=1$, which defines the $A_4$ singularity. In fact, the higher-order singularity points shown in Fig.~\ref{fig:phase_diagram} can be found with the guide of the $\lambda$ curves, as mentioned before. For GMCT with $N=3$, the shapes of the $\lambda$ curves are similar to MCT but quantitatively different. The peak of $\lambda(\theta)$ at $\delta=0.0465$ for $N=3$ is $~0.97\approx 1$, and hence we conjecture the $N=3$ $A_4$ point to exist for a $\delta$ value close to or slightly lower than $0.0465$. From the values of $\lambda$ when $\lambda<1$, we can also calculate the corresponding exponents $a$, $b$ and $\gamma$ by virtue of Eqs.\ (\ref{eq:ablambda}) and  (\ref{eq:gammaab}), i.e., the scaling laws of the relaxation dynamics near $A_2$ singularities. 
Notice that at large $\theta$, $\lambda$ must go to the value of the hard-sphere limit, i.e., $\lambda=0.73$ for $N=2$ (thin black line in Fig.~\ref{fig:lambda}) and $\lambda=0.78$ for $N=3$ ( thin blue line in Fig.~\ref{fig:lambda}), for any value of $\delta$. This puts a physical constraint on possible rescaling approaches; in particular,  for fixed $\delta$, the curves of $\lambda(\theta)$ for different closure levels $N$ cannot collapse to one curve by simply shifting $\theta$. This is a further manifestation of the non-trivial effects imposed by higher-order GMCT on the glassy dynamics of sticky hard spheres, consistent with the subtle differences in the phase diagrams of Sec.\ \ref{sec:phasediagram}.

In view of the above findings, let us also make a general remark on the rescaling procedure widely used in standard MCT. For sticky hard spheres, it is in a sense  accidental that a linear rescaling of only the packing fraction $\varphi$
and the reduced temperature $\theta$ at constant $\delta$ has been so successfully applied for the $A_3$ and $A_4$ singularities. The success of this rescaling approach relies on the fact that the shape of $\varphi^c(\theta)$ near the $A_3$ or $A_4$ singularities and the shape of the peaks of $\lambda(\theta)$ remain very similar under different closure levels $N$ for a given $\delta$. We expect this similarity to continue as $N$ increases until $N\rightarrow\infty$. It is this inherent robustness that renders a simple rescaling of $\theta$ and $\varphi$ in MCT sufficient to predict the $A_3$ and $A_4$ singularities within tolerance errors.\cite{sciortino2003evidence} However, regardless of how we rescale $\varphi$ and $\theta$, it is strictly impossible to merge the full $\varphi^c(\theta)$ and $\lambda(\theta)$ curves \textit{without simultaneously rescaling} $\delta$. Consequently, we argue that it is \textit{a priori} somewhat imprudent to test MCT by only rescaling two parameters; ideally one should first solve the first two or three levels of the GMCT hierarchy to establish a robust trend of the transitions before applying such a rescaling approach.


\subsection{Relaxation dynamics}

\begin{figure}[h!]
	\centering
	\includegraphics[width=7cm]{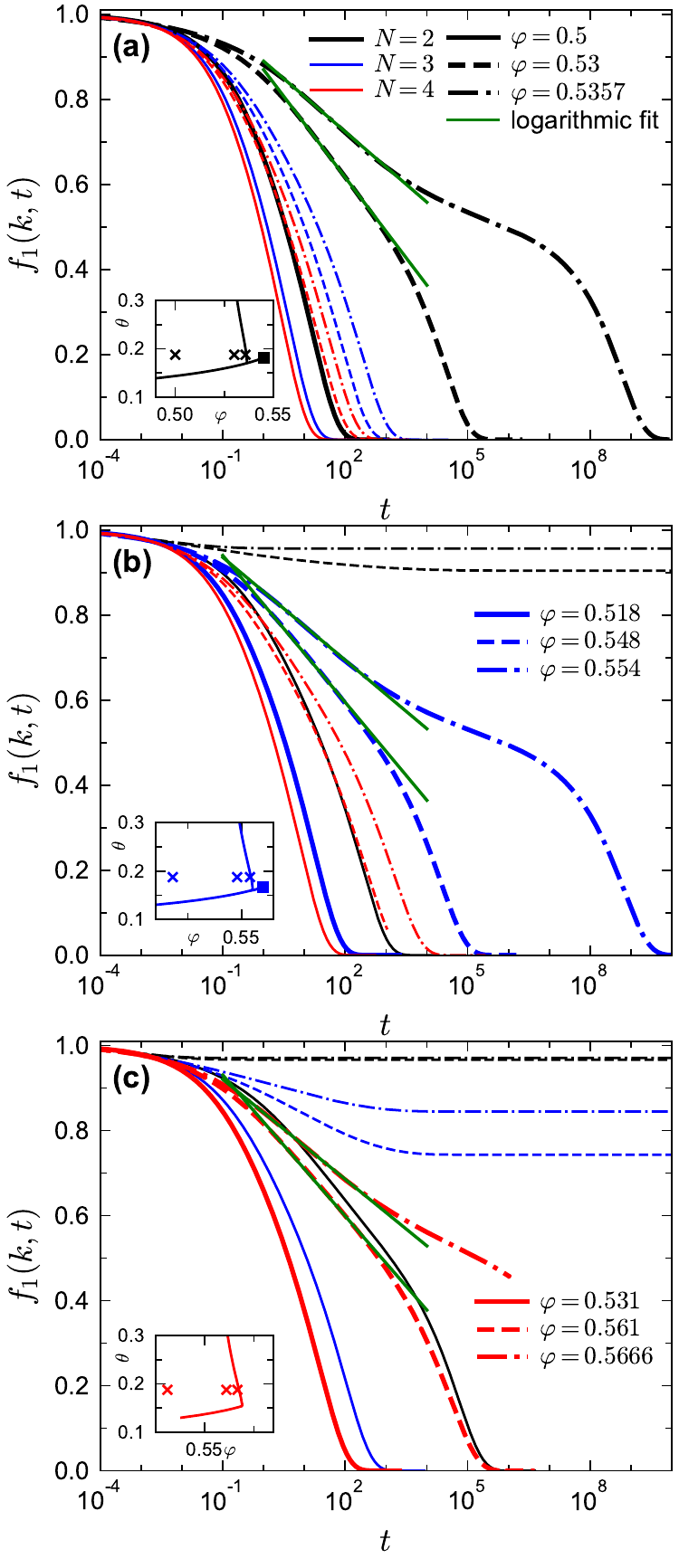}
	\caption{Normalized two-point density correlation functions $f_1(k,t)$ for liquid states at $\delta=0.03$ and $\theta=0.1875$ near the $A_3$ singularities  under different GMCT closure levels. The crosses in the inset indicate the considered liquid-state points. (a) Results for packing fractions $\varphi=0.5,\ 0.53,\ 0.5357$; these points lie near the $A_3$ singularity ($\varphi=0.545$,$\theta=0.182$) predicted from GMCT under closure level $N=2$. 
	The solid, dashed, and dash-dotted lines correspond to different packing fractions; different colors correspond to different MF closure levels.  The green lines are fits to the expression $f_1(k,t)=f^c_1(k,t)-C_1(k)\ln (t)$, where $C_1(k)$ is a wavenumber-dependent constant. (b) Same as in (a), but for packing fractions $\varphi=0.518,\ 0.548,\ 0.554$; these points lie near the $A_3$ singularity (($\varphi=0.56$,$\theta=0.167$) predicted from GMCT under closure level $N=3$. (c) Same as in (a), but for packing fractions $\varphi=0.531,\ 0.561,\ 0.5666$. It takes more than one month to obtain the data for the red dash-dotted line ($N=4$); we expect this curve to ultimately decay to $0$. }
	\label{fig:Fkt_A3}
\end{figure}

We now consider the time dependence of the dynamic density correlators. The effect of higher-order GMCT on the structural relaxation dynamics of monodisperse hard spheres has already been studied in detail.\cite{janssen2015microscopic,luo2020generalized1,luo2020generalized2} As already mentioned in Sec.\ \ref{sec:theory}, one of the conclusions of these recent studies is that for all packing fractions and wavenumbers, GMCT MF closures provide an upper bound to the dynamics while the exponential closures give a lower bound. The two types of closures systematically converge to each other when the closure level $N$ increases. More specifically, if we focus on the MF closures, the predicted relaxation becomes faster in a convergent manner. Here we find that this pattern also applies to sticky hard spheres, i.e., for a given state (fixed packing fraction, fixed reduced temperature, and fixed relative width of the potential well), the relaxation is faster when the MF closure level $N$ increases. We can clearly see this in Fig.~\ref{fig:Fkt_A3}, where the normalized intermediate scattering functions $f_1(k,t)=F_1(k,t)/S(k)$ at nine different liquid states under different MF closure levels are presented. For example, it can be seen that the $N=3$ results (blue lines in Fig.~\ref{fig:Fkt_A3}) always decay faster than the corresponding $N=2$ (black) curves, and more generally the differences between the curves for closure levels $N+1$ and $N$ become smaller as $N$ increases. These findings are thus consistent with a uniform convergent trend of the GMCT hierarchy. 

\begin{figure}[h!]
	\centering
	\includegraphics[width=8.3cm]{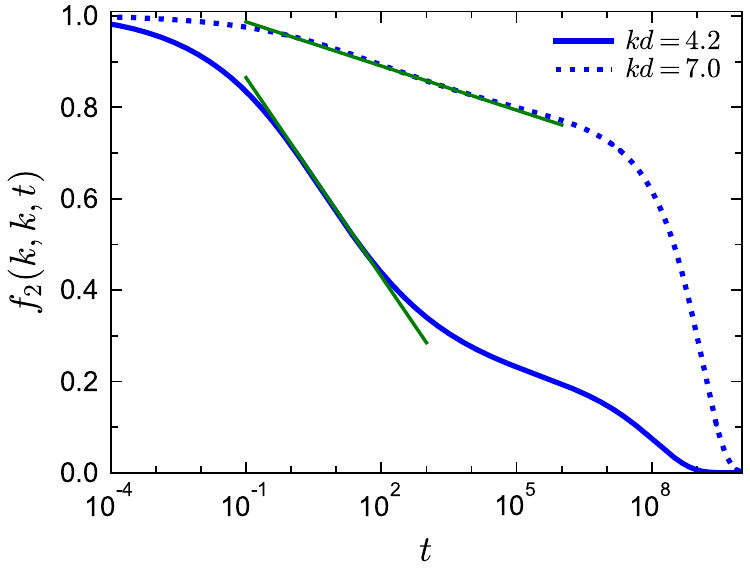}
	\caption{Normalized four-point density correlation functions $f_2(k,k,t)$ for liquid states at $\delta=0.03$, $\theta=0.1875$ and $\varphi=0.554$ (corresponding to the right cross in the inset of Fig.\ \ref{fig:Fkt_A3}(b)) near the $A_3$ singularities predicted from GMCT with closure level $N=3$. The solid and dotted curves correspond to wavenumber $k=4.2$ and $k=7.0$, respectively. The green lines are fits to the expression $f_2(k,k,t)=f^c_2(k,k)-C_2(k,k)\ln (t)$, where $C_2(k,k)$ is a wavenumber-dependent constant.}
	\label{fig:F2kt_A3}
\end{figure}

The more important properties of the dynamics are the scaling laws in the $\beta-$ and $\alpha-$relaxation regimes, since these are widely used for analyzing experimental and simulation data.
Within MCT, the scaling laws of Eqs.\ (\ref{eq:fachG}), (\ref{eq:GA}), (\ref{eq:GB}), and (\ref{eq:alphascaling}) are known to be applicable for liquid states that lie near the $A_2$ singularities 
but away from the $A_3$ and $A_4$ singularities.
Note that in this case the exponent parameter $\lambda$ must be smaller than 1, or more strictly $\lambda<0.9$, to ensure that the liquid state is sufficiently far away from the $A_3$ and $A_4$ points. 
We find that the above scaling laws are also preserved within GMCT under MF closures near the corresponding $A_2$ singularities, i.e., near both the GMCT-predicted liquid-to-repulsive-glass and liquid-to-attractive-glass transitions.  
This is natural and reasonable given the similar mathematical form of the MCT and GMCT equations. However, when the liquid states are close to the $A_3$ and $A_4$ singularities, the above scaling laws for $A_2$ break down and new behaviors of the relaxation dynamics will appear. Hence we will focus on the dynamics near these higher-order singularities in the following. 

We first study the dynamics near the $A_3$ singularities for different GMCT closure levels. The $A_3$ points at $\delta=0.03$ are $(\theta\approx0.182,\varphi\approx0.545)$ and $(\theta\approx0.167,\varphi\approx0.56)$ for $N=2$ and $N=3$, respectively. Even though we could not accurately determine the location of the $A_3$ point for $N=4$ due to limited computing power, we are certain this point must exist and should lie at a $\varphi$ slightly above $0.57$ and a $\theta$ slightly above $0.156$ for $\delta=0.03$, since there is an $N=4$ glass-glass transition found at $\varphi=0.57$ and $\theta=0.156$ for $\delta=0.03$ (see also the non-ergodicity parameters in Fig.\ \ref{fig:fc_vs_theta}).
For our $A_3$ analysis we fix $\theta=0.1875$ and select three liquid states with different packing fractions for each closure level $N$; this value of $\theta$ is close to the $\theta$ of the $A_3$ singularity for all considered $N$. The results for the predicted normalized dynamical two-point density correlation functions $f_1(k,t)$ are shown in Fig.\ \ref{fig:Fkt_A3}. 
As we get closer to the $A_3$ singularities, we find that the relaxation of $f_1(k,t)$ becomes more stretched compared to the power laws described in Eqs.\ (\ref{eq:GA}) and (\ref{eq:GB}), and instead a logarithmic relaxation regime appears. This logarithmic decay of $f_1(k,t)$ for liquid states near $A_3$ singularities, described via Eq.~(\ref{eq:Gtlog}), is applicable for all closure levels $N$ (see the green dashed fitted curves in Fig.~\ref{fig:Fkt_A3}). 
The logarithmic law for $f_1(k,t)$ near $A_3$ singularities within GMCT qualitatively agrees with the one in MCT, and hence our GMCT results also agree with the body of simulation and experimental data that have already been qualitatively compared to MCT.\cite{zaccarelli2002confirmation,sciortino2003evidence,pham2002multiple} More generally, we find that higher-order GMCT inherits all the main qualitative features of the MCT dynamics near $A_3$ singularities. In particular, upon approaching the $A_3$ point from the liquid side (black, blue, and red dashed lines in Fig.\ \ref{fig:Fkt_A3}(a), (b), and (c), respectively), we can also see that the $\alpha$-relaxation process does not start with von Schweidler law of Eq.\ (\ref{eq:GB}), and as a consequence the superposition principle of Eq.\ (\ref{eq:alphascaling}) is disobeyed.\cite{dawson2000higher} The only main difference between MCT and GMCT is that the corresponding packing fractions are improved, since the $A_3$ point becomes more accurate upon increasing the closure level. 

 Besides the logarithmic decay for $f_1(k,t)$, we find that this logarithmic decay is also applicable for $2n$-point density correlation functions with $n>1$. For example, in Fig.\ \ref{fig:F2kt_A3} we show $f_2(k,k,t)$ near the $A_3$ singularity under closure level $N=3$ for two different wavenumbers. The fit with $f_2(k,k,t)=f^c_2(k,k)-C_2(k,k)\ln (t)$, where $C_2(k,k)$ is a wavenumber-dependent constant, is applicable for more than three decades in time. To the best of our knowledge, no empirical results for $f_2(k,k,t)$ or similar higher-order dynamical density correlation functions have been reported in this glassy regime. We argue that the logarithmic decay for higher-order correlators can be another strict test for both GMCT and the $A_3$ singularities predicted from it for sticky hard spheres.

\begin{figure}[h!]
	\centering
	\includegraphics[width=8.3cm]{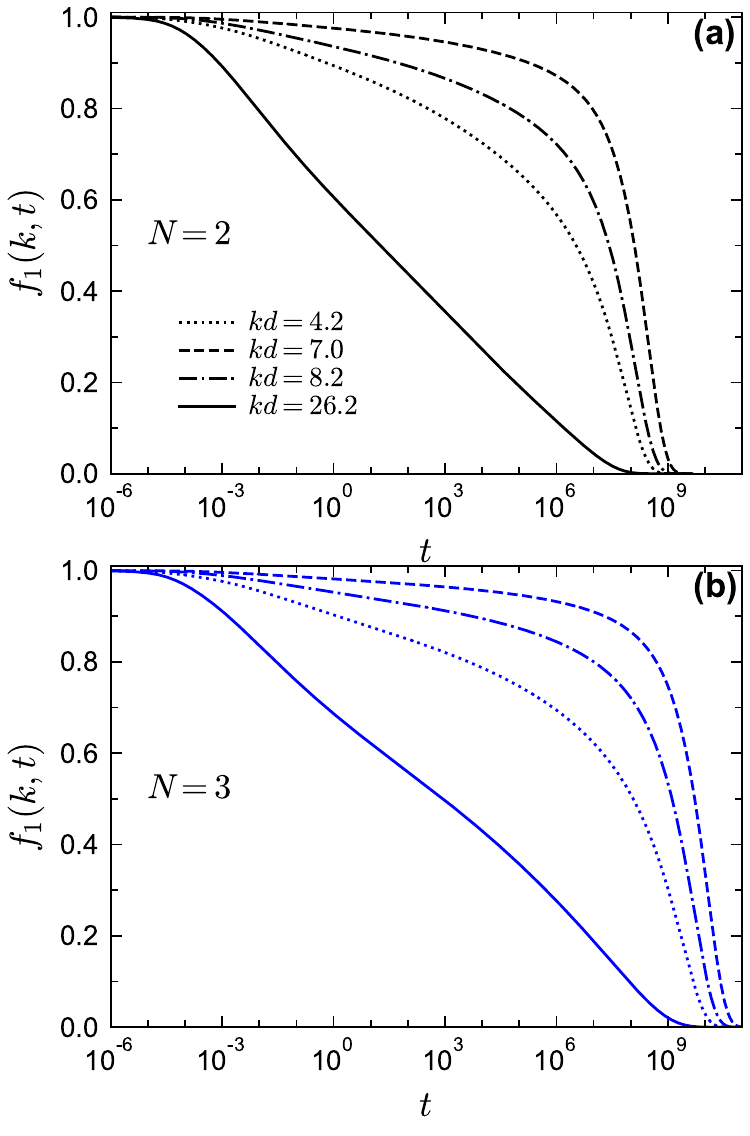}
	\caption{Normalized two-point density correlation functions $f_1(k,t)$ at different wavenumbers near the $A_4$ singularities under different GMCT closure levels. (a) $f_1(k,t)$ under closure level $N=2$ at $\varphi=0.524$ and $\theta=0.234$, near the $A_4$ point ($\varphi=0.526$, $\theta=0.234$) predicted from MCT. (b) $f_1(k,t)$ under closure level $N=3$ at $\varphi=0.543$ and $\theta=0.22$ near the $A_4$ point ($\varphi=0.544$, $\theta=0.22$) predicted from GMCT under closure level $N=3$. }
	\label{fig:Fkt_A4}
\end{figure}

\begin{figure}[h!]
	\centering
	\includegraphics[width=8.3cm]{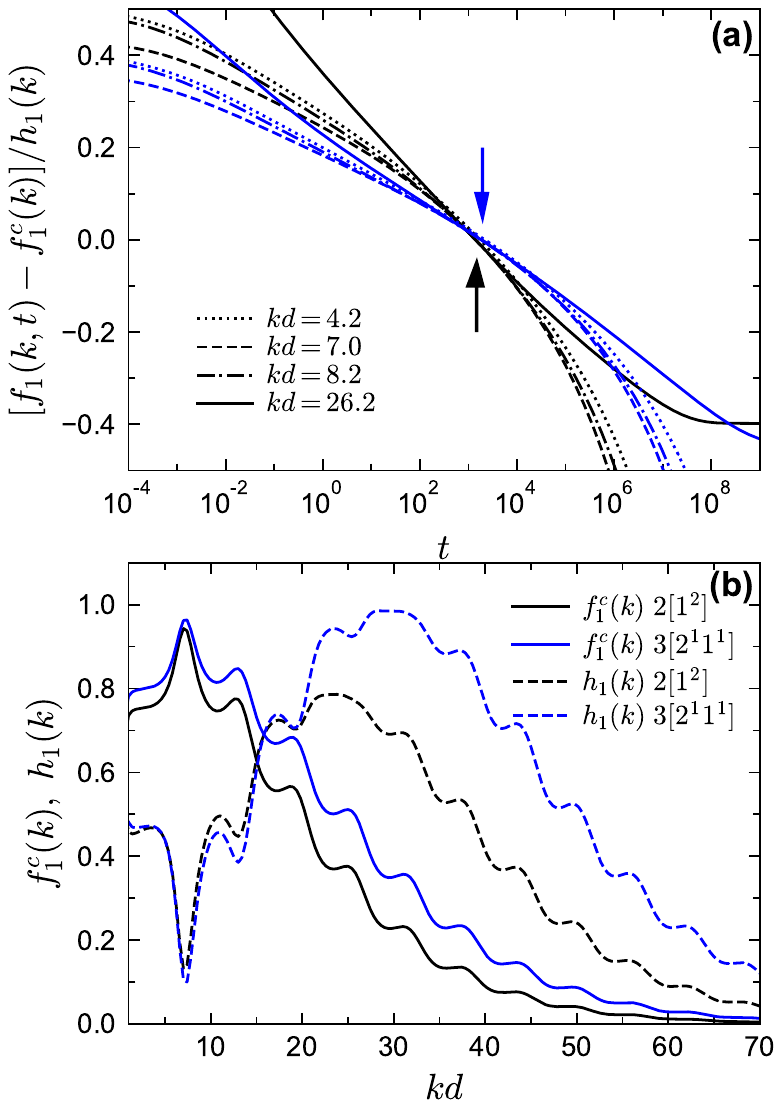}
	\caption{Scaling laws near the $A_4$ singularity point. (a) Rescaled two-point density correlation functions $\left[f_1(k,t)-f_1^c(k)\right]/h_1(k)$ at the same liquid states as in Fig.\ \ref{fig:Fkt_A4}, i.e., near the corresponding $A_4$ points for GMCT closure levels $N=2$ (black lines) and $N=3$ (blue lines). The black (blue) arrow indicates the position of $G(t_-)=0$ with $t_-=1.48\times 10^3$ ($t_-=1.95\times 10^3$) for closure level $N=2$ ($N=3$).
	(b) The functions $f_1^c(k)$ and $h_1(k)$ obtained from the corresponding $A_4$ singularities using Eq.\ \ref{eq:GMCT_long} and Eq.\ \ref{eq:hn}. }
	\label{fig:Fkt_A4_scaling}
\end{figure}

Finally let us go to the dynamics of the density correlation functions near the $A_4$ singularities. The regime of the very stretched and nearly logarithmic decay of $f_1(k,t)$ is found for both closure levels $N=2$ and $N=3$, as indicated by the solid lines in Fig.\ \ref{fig:Fkt_A4}.  The range of this regime can be much wider than the range near $A_3$ in Fig.\ \ref{fig:Fkt_A3}, although for $N=3$ [blue solid line in Fig.\ \ref{fig:Fkt_A4}(b)] the range of the strictly logarithmic decay is not as wide as the one for $N=2$ [black solid line in Fig.\ \ref{fig:Fkt_A4}(a)]. We attribute this to the lower numerical accuracy of the $A_4$ singularity for $N=3$ (with a maximum $\lambda=0.97$ at $\delta=0.0465$ instead of $\lambda=1$). We have also verified that the logarithmic decay is found for $f_2(k_1,k_2,t)$ at several wavenumbers (not shown here). We further test the time-wavenumber factorization property, Eq.\ (\ref{eq:fachG}), near the $A_4$ singularities. If the factorization property holds, the rescaled correlation functions $\left[f_n(k_1,\hdots,k_n,t)-f_n^c(k_1,\hdots,k_n)\right]/h_n(k_1,\hdots,k_n)$ should collapse on the common function $G(t)$. In particular, all correlators should cross their plateau value $f^c_n(k_1,\hdots,k_n,t)$ at the same time $t_-$ with $G(t_-)=0$. We show the rescaled correlation function of $f_1(k,t)$ in Fig.\ \ref{fig:Fkt_A4_scaling}(a) with the $f^c_1(k)$ and $h_1(k)$ [shown in Fig.\ \ref{fig:Fkt_A4_scaling}(b)] from the $A_4$ singularity point under the corresponding closure level $N$. We find that for both $N=2$ and $N=3$, the factorization property is still satisfied (see Fig.\ \ref{fig:Fkt_A4_scaling}(a) with the arrows indicating the $t_-$). However, the validity range is only around two decades in time, which is much narrower than the one near $A_2$ singularities, for which the factorization property applies over approximately four decades.\cite{luo2020generalized1,dawson2000higher}


\section{Conclusion}
In this work, we have presented a detailed numerical analysis of the glassy dynamics of sticky hard spheres using first-principles-based microscopic generalized mode-coupling theory. We have obtained the phase diagrams in the three-dimensional control-parameter space under different mean-field closure levels in GMCT. Upon increasing the closure level $N$ of the GMCT hierarchy, i.e., by expanding the dynamical equations in terms of increasingly higher order density correlations, the phase diagrams become closer to the empirical results. We have clarified that the improvement from GMCT is not merely a simple rescaling of control parameters, as is usually done in standard MCT analyses; rather, higher-order GMCT yields subtle but intrinsically  different results compared to MCT, as reflected in  the shifted phase diagrams of sticky hard spheres relative to the hard-sphere case, and the exponent parameter $\lambda$. 

From the phase diagrams and the non-ergodicity parameters, we have demonstrated that the peculiar features of dense sticky hard spheres, i.e., glassy reentrance, a glass-glass transition, and higher-order $A_3$ and $A_4$ singularities, are also successfully predicted from GMCT. The positions of these transition points are also more accurate than those obtained from standard MCT. These triumphs indicate that the complex interplay of repulsive and attractive short-ranged interactions can be adequately captured within GMCT. We have also demonstrated that when the attraction is very strong, the difference of the predicted critical packing fractions at any given width of the potential well between GMCT and MCT becomes much larger. This implies that, compared to repulsive interactions, the attractive part of the interaction potential may play a more important role in the higher-order density correlation functions. This non-trivial effect of attractions has also been reflected in the different patterns of the non-ergodicity parameters of the four-point density correlation functions $f^c_2(k_1,k_2)$ for the repulsive-glass and attractive-glass phase. We hypothesize that these higher-order density correlations may also contain new, and non-trivial, information on dynamical heterogeneity,\cite{reichman2005comparison} but further research is needed to clarify their precise physical interpretation. 

Within higher-order GMCT, we have shown that the celebrated scaling laws for $A_2$ singularities are no longer applicable when the considered state point is in the vicinity of an $A_3$ or $A_4$ singularity. In that case, we find a logarithmic decay in the dynamics that is consistent with the MCT form of Eq.\ (\ref{eq:Gtlog}). We emphasize, however, that this logarithmic decay is applicable not only for the conventionally studied intermediate scattering function $f_1(k,t)$, but also for higher-order density correlation functions such as $f_2(k,k,t)$. The latter of course needs to be further verified in simulations or experiments.

From all the results above, we conclude that GMCT can make detailed first-principles-based predictions of the microscopic dynamics of sticky hard spheres well beyond the standard-MCT regime, whilst maintaining many of the qualitative features of MCT. 
Interestingly, although a significant body of experimental\cite{mallamace2000kinetic,eckert2002re,pham2002multiple,pham2004glasses,kaufman2006direct} and simulation\cite{puertas2002comparative,zaccarelli2002confirmation,foffi2002evidence,zaccarelli2003activated,sciortino2003evidence,zaccarelli2004numerical,foffi2004dynamical,saika2004effect,reichman2005comparison,moreno2006anomalous} work has corroborated the qualitative MCT scenario for sticky hard spheres,\cite{puertas2002comparative,zaccarelli2002confirmation,foffi2002evidence,zaccarelli2003activated,sciortino2003evidence,zaccarelli2004numerical,foffi2004dynamical} a recent swap Monte Carlo simulation\cite{fullerton2020glassy} disputes the existence of the 
MCT-predicted liquid-glass transition, the glass-glass transition, and all the higher-order $A_l$ singularities, claiming that these MCT-predicted phenomena in fact lie within a unique ergodic region. 
Our results support this conclusion in the sense that the MCT-predicted glass transitions are shifted toward higher packing fractions within higher-order GMCT, and as $N$ increases they might become close to or even higher than the highest packing fractions accessible in simulation.
However, we here argue that the liquid-glass transition, glass-glass transition and higher-order singularities could still exist, possibly at even higher packing fractions than those probed by state-of-the-art simulations.\cite{fullerton2020glassy} 
We also emphasize that our theoretical results are based on a monodisperse sticky hard sphere model, while the swap Monte Carlo simulations of Ref.\ \cite{fullerton2020glassy} assume a relatively large polydispersity of $23\%$. It is known in general that polydispersity can significantly affect the glassy dynamics and the corresponding phase diagrams,\cite{ciarella2021multi} and hence this  aspect should be also carefully studied in future work.
Finally, the high-order singularities and logarithmic relaxation of sticky hard spheres may also have a deep analogy to the glassy behavior of randomly pinned supercooled liquids and that of the random field Ising model, as discovered within the inhomogeneous MCT. \cite{nandi2014critical} We believe this analogy would also be preserved in our GMCT framework, although further work on this is needed.

Since the sticky hard sphere model contains both attractive and repulsive interactions, we may expect that GMCT is also a suitable framework for other systems with short-ranged attractive interactions. It remains to be explored, however, whether the current GMCT framework can ultimately provide a universal description of the elusive structure-dynamics link in glass-forming materials. 
Firstly, there are still some subtle approximations in our current theory, such as the diagonal approximation for the dynamical higher-order density correlation functions in the GMCT memory functions,\cite{SimonePhdthesis} and the Gaussian factorization and convolution approximation for all higher-order \textit{static} density correlation functions.\cite{szamel2003colloidal,janssen2015microscopic} Recent studies suggest that higher-order and local structural metrics also encode rich physics and contribute to the dynamics in a non-trivial manner,\cite{liu2019machine,zhang2020revealing,robinson2019morphometric} but these structural quantities are all simply regarded as being implicitly included in $S(k)$ or neglected in the current GMCT framework. Future work will be done to test how accurate these approximations are and how one might admit more detailed structural information into the theory. Second, until now, GMCT has been tested only for systems with purely repulsive or repulsive with short-ranged attractive isotropic potentials. These systems all belong to the class of fragile glass formers.\cite{scopigno2003fragility,martinez2001thermodynamic} More work needs be done to test GMCT for strong glass formers such as silica \cite{sciortino2001debye}, or even superstrong systems such as vitrimers.\cite{ciarella2019understanding} 

\section*{Conflicts of interest}
There are no conflicts to declare.

\section*{Acknowledgements}
We acknowledge the Netherlands Organisation for Scientific Research (NWO) for financial support through a START-UP grant.





\bibliography{sticky_hard_sphere} 

\providecommand*{\mcitethebibliography}{\thebibliography}
\csname @ifundefined\endcsname{endmcitethebibliography}
{\let\endmcitethebibliography\endthebibliography}{}
\begin{mcitethebibliography}{75}
\providecommand*{\natexlab}[1]{#1}
\providecommand*{\mciteSetBstSublistMode}[1]{}
\providecommand*{\mciteSetBstMaxWidthForm}[2]{}
\providecommand*{\mciteBstWouldAddEndPuncttrue}
  {\def\EndOfBibitem{\unskip.}}
\providecommand*{\mciteBstWouldAddEndPunctfalse}
  {\let\EndOfBibitem\relax}
\providecommand*{\mciteSetBstMidEndSepPunct}[3]{}
\providecommand*{\mciteSetBstSublistLabelBeginEnd}[3]{}
\providecommand*{\EndOfBibitem}{}
\mciteSetBstSublistMode{f}
\mciteSetBstMaxWidthForm{subitem}
{(\emph{\alph{mcitesubitemcount}})}
\mciteSetBstSublistLabelBeginEnd{\mcitemaxwidthsubitemform\space}
{\relax}{\relax}

\bibitem[Debenedetti and Stillinger(2001)]{debenedetti2001supercooled}
P.~G. Debenedetti and F.~H. Stillinger, \emph{Nature}, 2001, \textbf{410},
  259\relax
\mciteBstWouldAddEndPuncttrue
\mciteSetBstMidEndSepPunct{\mcitedefaultmidpunct}
{\mcitedefaultendpunct}{\mcitedefaultseppunct}\relax
\EndOfBibitem
\bibitem[Berthier and Biroli(2011)]{berthier2011theoretical}
L.~Berthier and G.~Biroli, \emph{Reviews of Modern Physics}, 2011, \textbf{83},
  587\relax
\mciteBstWouldAddEndPuncttrue
\mciteSetBstMidEndSepPunct{\mcitedefaultmidpunct}
{\mcitedefaultendpunct}{\mcitedefaultseppunct}\relax
\EndOfBibitem
\bibitem[G{\"o}tze(2009)]{gotze2008complex}
W.~G{\"o}tze, \emph{Complex dynamics of glass-forming liquids: A Mode-Coupling
  Theory}, Oxford University Press, Oxford, 2009\relax
\mciteBstWouldAddEndPuncttrue
\mciteSetBstMidEndSepPunct{\mcitedefaultmidpunct}
{\mcitedefaultendpunct}{\mcitedefaultseppunct}\relax
\EndOfBibitem
\bibitem[Tarjus \emph{et~al.}(2005)Tarjus, Kivelson, Nussinov, and
  Viot]{tarjus2005frustration}
G.~Tarjus, S.~A. Kivelson, Z.~Nussinov and P.~Viot, \emph{Journal of Physics:
  Condensed Matter}, 2005, \textbf{17}, R1143\relax
\mciteBstWouldAddEndPuncttrue
\mciteSetBstMidEndSepPunct{\mcitedefaultmidpunct}
{\mcitedefaultendpunct}{\mcitedefaultseppunct}\relax
\EndOfBibitem
\bibitem[Kirkpatrick and Thirumalai(2015)]{kirkpatrick2015colloquium}
T.~Kirkpatrick and D.~Thirumalai, \emph{Reviews of Modern Physics}, 2015,
  \textbf{87}, 183\relax
\mciteBstWouldAddEndPuncttrue
\mciteSetBstMidEndSepPunct{\mcitedefaultmidpunct}
{\mcitedefaultendpunct}{\mcitedefaultseppunct}\relax
\EndOfBibitem
\bibitem[Liu \emph{et~al.}(2019)Liu, Fu, Yang, Xu, and Bauchy]{liu2019machine}
H.~Liu, Z.~Fu, K.~Yang, X.~Xu and M.~Bauchy, \emph{Journal of Non-Crystalline
  Solids}, 2019,  119419\relax
\mciteBstWouldAddEndPuncttrue
\mciteSetBstMidEndSepPunct{\mcitedefaultmidpunct}
{\mcitedefaultendpunct}{\mcitedefaultseppunct}\relax
\EndOfBibitem
\bibitem[Ma \emph{et~al.}(2019)Ma, Davidson, Still, Ivancic, Schoenholz, Liu,
  and Yodh]{ma2019heterogeneous}
X.~Ma, Z.~S. Davidson, T.~Still, R.~J. Ivancic, S.~Schoenholz, A.~Liu and
  A.~Yodh, \emph{Physical Review Letters}, 2019, \textbf{122}, 028001\relax
\mciteBstWouldAddEndPuncttrue
\mciteSetBstMidEndSepPunct{\mcitedefaultmidpunct}
{\mcitedefaultendpunct}{\mcitedefaultseppunct}\relax
\EndOfBibitem
\bibitem[Schoenholz \emph{et~al.}(2016)Schoenholz, Cubuk, Sussman, Kaxiras, and
  Liu]{schoenholz2016structural}
S.~S. Schoenholz, E.~D. Cubuk, D.~M. Sussman, E.~Kaxiras and A.~J. Liu,
  \emph{Nature Physics}, 2016, \textbf{12}, 469--471\relax
\mciteBstWouldAddEndPuncttrue
\mciteSetBstMidEndSepPunct{\mcitedefaultmidpunct}
{\mcitedefaultendpunct}{\mcitedefaultseppunct}\relax
\EndOfBibitem
\bibitem[Bapst \emph{et~al.}(2020)Bapst, Keck, Grabska-Barwi{\'n}ska, Donner,
  Cubuk, Schoenholz, Obika, Nelson, Back,
  Hassabis,\emph{et~al.}]{bapst2020unveiling}
V.~Bapst, T.~Keck, A.~Grabska-Barwi{\'n}ska, C.~Donner, E.~D. Cubuk, S.~S.
  Schoenholz, A.~Obika, A.~W. Nelson, T.~Back, D.~Hassabis \emph{et~al.},
  \emph{Nature Physics}, 2020, \textbf{16}, 448--454\relax
\mciteBstWouldAddEndPuncttrue
\mciteSetBstMidEndSepPunct{\mcitedefaultmidpunct}
{\mcitedefaultendpunct}{\mcitedefaultseppunct}\relax
\EndOfBibitem
\bibitem[Bengtzelius \emph{et~al.}(1984)Bengtzelius, G{\"o}tze, and
  Sjolander]{bengtzelius1984dynamics}
U.~Bengtzelius, W.~G{\"o}tze and A.~Sjolander, \emph{Journal of Physics C:
  Solid State Physics}, 1984, \textbf{17}, 5915\relax
\mciteBstWouldAddEndPuncttrue
\mciteSetBstMidEndSepPunct{\mcitedefaultmidpunct}
{\mcitedefaultendpunct}{\mcitedefaultseppunct}\relax
\EndOfBibitem
\bibitem[Leutheusser(1984)]{leutheusser1984dynamical}
E.~Leutheusser, \emph{Physical Review A}, 1984, \textbf{29}, 2765\relax
\mciteBstWouldAddEndPuncttrue
\mciteSetBstMidEndSepPunct{\mcitedefaultmidpunct}
{\mcitedefaultendpunct}{\mcitedefaultseppunct}\relax
\EndOfBibitem
\bibitem[Reichman and Charbonneau(2005)]{reichman2005mode}
D.~R. Reichman and P.~Charbonneau, \emph{Journal of Statistical Mechanics:
  Theory and Experiment}, 2005, \textbf{2005}, P05013\relax
\mciteBstWouldAddEndPuncttrue
\mciteSetBstMidEndSepPunct{\mcitedefaultmidpunct}
{\mcitedefaultendpunct}{\mcitedefaultseppunct}\relax
\EndOfBibitem
\bibitem[Janssen(2018)]{janssen2018mode}
L.~M.~C. Janssen, \emph{Frontiers in Physics}, 2018, \textbf{6}, 97\relax
\mciteBstWouldAddEndPuncttrue
\mciteSetBstMidEndSepPunct{\mcitedefaultmidpunct}
{\mcitedefaultendpunct}{\mcitedefaultseppunct}\relax
\EndOfBibitem
\bibitem[Kob \emph{et~al.}(1997)Kob, Donati, Plimpton, Poole, and
  Glotzer]{kob1997dynamical}
W.~Kob, C.~Donati, S.~J. Plimpton, P.~H. Poole and S.~C. Glotzer,
  \emph{Physical Review Letters}, 1997, \textbf{79}, 2827\relax
\mciteBstWouldAddEndPuncttrue
\mciteSetBstMidEndSepPunct{\mcitedefaultmidpunct}
{\mcitedefaultendpunct}{\mcitedefaultseppunct}\relax
\EndOfBibitem
\bibitem[Glotzer \emph{et~al.}(2000)Glotzer, Novikov, and
  Schr{\o}der]{glotzer2000time}
S.~C. Glotzer, V.~N. Novikov and T.~B. Schr{\o}der, \emph{The Journal of
  Chemical Physics}, 2000, \textbf{112}, 509--512\relax
\mciteBstWouldAddEndPuncttrue
\mciteSetBstMidEndSepPunct{\mcitedefaultmidpunct}
{\mcitedefaultendpunct}{\mcitedefaultseppunct}\relax
\EndOfBibitem
\bibitem[La{\v{c}}evi{\'c} \emph{et~al.}(2003)La{\v{c}}evi{\'c}, Starr,
  Schr{\o}der, and Glotzer]{lavcevic2003spatially}
N.~La{\v{c}}evi{\'c}, F.~W. Starr, T.~Schr{\o}der and S.~Glotzer, \emph{The
  Journal of chemical physics}, 2003, \textbf{119}, 7372--7387\relax
\mciteBstWouldAddEndPuncttrue
\mciteSetBstMidEndSepPunct{\mcitedefaultmidpunct}
{\mcitedefaultendpunct}{\mcitedefaultseppunct}\relax
\EndOfBibitem
\bibitem[Zhang and Kob(2020)]{zhang2020revealing}
Z.~Zhang and W.~Kob, \emph{Proceedings of the National Academy of Sciences},
  2020, \textbf{117}, 14032--14037\relax
\mciteBstWouldAddEndPuncttrue
\mciteSetBstMidEndSepPunct{\mcitedefaultmidpunct}
{\mcitedefaultendpunct}{\mcitedefaultseppunct}\relax
\EndOfBibitem
\bibitem[Szamel(2003)]{szamel2003colloidal}
G.~Szamel, \emph{Physical Review Letters}, 2003, \textbf{90}, 228301\relax
\mciteBstWouldAddEndPuncttrue
\mciteSetBstMidEndSepPunct{\mcitedefaultmidpunct}
{\mcitedefaultendpunct}{\mcitedefaultseppunct}\relax
\EndOfBibitem
\bibitem[Wu and Cao(2005)]{wu2005high}
J.~Wu and J.~Cao, \emph{Physical Review Letters}, 2005, \textbf{95},
  078301\relax
\mciteBstWouldAddEndPuncttrue
\mciteSetBstMidEndSepPunct{\mcitedefaultmidpunct}
{\mcitedefaultendpunct}{\mcitedefaultseppunct}\relax
\EndOfBibitem
\bibitem[Janssen and Reichman(2015)]{janssen2015microscopic}
L.~M.~C. Janssen and D.~R. Reichman, \emph{Physical Review Letters}, 2015,
  \textbf{115}, 205701\relax
\mciteBstWouldAddEndPuncttrue
\mciteSetBstMidEndSepPunct{\mcitedefaultmidpunct}
{\mcitedefaultendpunct}{\mcitedefaultseppunct}\relax
\EndOfBibitem
\bibitem[Luo and Janssen(2020)]{luo2020generalized1}
C.~Luo and L.~M.~C. Janssen, \emph{The Journal of Chemical Physics}, 2020,
  \textbf{153}, 214507\relax
\mciteBstWouldAddEndPuncttrue
\mciteSetBstMidEndSepPunct{\mcitedefaultmidpunct}
{\mcitedefaultendpunct}{\mcitedefaultseppunct}\relax
\EndOfBibitem
\bibitem[Luo and Janssen(2020)]{luo2020generalized2}
C.~Luo and L.~M.~C. Janssen, \emph{The Journal of Chemical Physics}, 2020,
  \textbf{153}, 214506\relax
\mciteBstWouldAddEndPuncttrue
\mciteSetBstMidEndSepPunct{\mcitedefaultmidpunct}
{\mcitedefaultendpunct}{\mcitedefaultseppunct}\relax
\EndOfBibitem
\bibitem[Kob and Andersen(1995)]{kob1995testing}
W.~Kob and H.~C. Andersen, \emph{Physical Review E}, 1995, \textbf{52},
  4134\relax
\mciteBstWouldAddEndPuncttrue
\mciteSetBstMidEndSepPunct{\mcitedefaultmidpunct}
{\mcitedefaultendpunct}{\mcitedefaultseppunct}\relax
\EndOfBibitem
\bibitem[Berthier and Tarjus(2010)]{berthier2010critical}
L.~Berthier and G.~Tarjus, \emph{Physical Review E}, 2010, \textbf{82},
  031502\relax
\mciteBstWouldAddEndPuncttrue
\mciteSetBstMidEndSepPunct{\mcitedefaultmidpunct}
{\mcitedefaultendpunct}{\mcitedefaultseppunct}\relax
\EndOfBibitem
\bibitem[Z{\"o}llmer \emph{et~al.}(2003)Z{\"o}llmer, R{\"a}tzke, Faupel, and
  Meyer]{zollmer2003diffusion}
V.~Z{\"o}llmer, K.~R{\"a}tzke, F.~Faupel and A.~Meyer, \emph{Physical Review
  Letters}, 2003, \textbf{90}, 195502\relax
\mciteBstWouldAddEndPuncttrue
\mciteSetBstMidEndSepPunct{\mcitedefaultmidpunct}
{\mcitedefaultendpunct}{\mcitedefaultseppunct}\relax
\EndOfBibitem
\bibitem[Nowak \emph{et~al.}(2017)Nowak, Holland-Moritz, Yang, Voigtmann,
  Kordel, Hansen, and Meyer]{nowak2017partial}
B.~Nowak, D.~Holland-Moritz, F.~Yang, T.~Voigtmann, T.~Kordel, T.~Hansen and
  A.~Meyer, \emph{Physical Review Materials}, 2017, \textbf{1}, 025603\relax
\mciteBstWouldAddEndPuncttrue
\mciteSetBstMidEndSepPunct{\mcitedefaultmidpunct}
{\mcitedefaultendpunct}{\mcitedefaultseppunct}\relax
\EndOfBibitem
\bibitem[Szamel and L{\"o}wen(1991)]{szamel1991mode}
G.~Szamel and H.~L{\"o}wen, \emph{Physical Review A}, 1991, \textbf{44},
  8215\relax
\mciteBstWouldAddEndPuncttrue
\mciteSetBstMidEndSepPunct{\mcitedefaultmidpunct}
{\mcitedefaultendpunct}{\mcitedefaultseppunct}\relax
\EndOfBibitem
\bibitem[Siebenb{\"u}rger \emph{et~al.}(2009)Siebenb{\"u}rger, Fuchs, Winter,
  and Ballauff]{siebenburger2009viscoelasticity}
M.~Siebenb{\"u}rger, M.~Fuchs, H.~Winter and M.~Ballauff, \emph{Journal of
  Rheology}, 2009, \textbf{53}, 707--726\relax
\mciteBstWouldAddEndPuncttrue
\mciteSetBstMidEndSepPunct{\mcitedefaultmidpunct}
{\mcitedefaultendpunct}{\mcitedefaultseppunct}\relax
\EndOfBibitem
\bibitem[Schrack and Franosch(2020)]{schrack2020mode}
L.~Schrack and T.~Franosch, \emph{Philosophical Magazine}, 2020, \textbf{100},
  1032--1057\relax
\mciteBstWouldAddEndPuncttrue
\mciteSetBstMidEndSepPunct{\mcitedefaultmidpunct}
{\mcitedefaultendpunct}{\mcitedefaultseppunct}\relax
\EndOfBibitem
\bibitem[Farago \emph{et~al.}(2012)Farago, Semenov, Meyer, Wittmer, Johner, and
  Baschnagel]{farago2012mode1}
J.~Farago, A.~Semenov, H.~Meyer, J.~Wittmer, A.~Johner and J.~Baschnagel,
  \emph{Physical Review E}, 2012, \textbf{85}, 051806\relax
\mciteBstWouldAddEndPuncttrue
\mciteSetBstMidEndSepPunct{\mcitedefaultmidpunct}
{\mcitedefaultendpunct}{\mcitedefaultseppunct}\relax
\EndOfBibitem
\bibitem[Farago \emph{et~al.}(2012)Farago, Meyer, Baschnagel, and
  Semenov]{farago2012mode2}
J.~Farago, H.~Meyer, J.~Baschnagel and A.~Semenov, \emph{Physical Review E},
  2012, \textbf{85}, 051807\relax
\mciteBstWouldAddEndPuncttrue
\mciteSetBstMidEndSepPunct{\mcitedefaultmidpunct}
{\mcitedefaultendpunct}{\mcitedefaultseppunct}\relax
\EndOfBibitem
\bibitem[Ciarella \emph{et~al.}(2019)Ciarella, Biezemans, and
  Janssen]{ciarella2019understanding}
S.~Ciarella, R.~A. Biezemans and L.~M.~C. Janssen, \emph{Proceedings of the
  National Academy of Sciences}, 2019, \textbf{116}, 25013--25022\relax
\mciteBstWouldAddEndPuncttrue
\mciteSetBstMidEndSepPunct{\mcitedefaultmidpunct}
{\mcitedefaultendpunct}{\mcitedefaultseppunct}\relax
\EndOfBibitem
\bibitem[Ruscher \emph{et~al.}(2020)Ruscher, Ciarella, Luo, Janssen, Farago,
  and Baschnagel]{ruscher2020glassy}
C.~Ruscher, S.~Ciarella, C.~Luo, L.~M.~C. Janssen, J.~Farago and J.~Baschnagel,
  \emph{Journal of Physics: Condensed Matter}, 2020, \textbf{33}, 064001\relax
\mciteBstWouldAddEndPuncttrue
\mciteSetBstMidEndSepPunct{\mcitedefaultmidpunct}
{\mcitedefaultendpunct}{\mcitedefaultseppunct}\relax
\EndOfBibitem
\bibitem[Puertas \emph{et~al.}(2002)Puertas, Fuchs, and
  Cates]{puertas2002comparative}
A.~M. Puertas, M.~Fuchs and M.~E. Cates, \emph{Physical Review Letters}, 2002,
  \textbf{88}, 098301\relax
\mciteBstWouldAddEndPuncttrue
\mciteSetBstMidEndSepPunct{\mcitedefaultmidpunct}
{\mcitedefaultendpunct}{\mcitedefaultseppunct}\relax
\EndOfBibitem
\bibitem[Zaccarelli \emph{et~al.}(2002)Zaccarelli, Foffi, Dawson, Buldyrev,
  Sciortino, and Tartaglia]{zaccarelli2002confirmation}
E.~Zaccarelli, G.~Foffi, K.~A. Dawson, S.~Buldyrev, F.~Sciortino and
  P.~Tartaglia, \emph{Physical Review E}, 2002, \textbf{66}, 041402\relax
\mciteBstWouldAddEndPuncttrue
\mciteSetBstMidEndSepPunct{\mcitedefaultmidpunct}
{\mcitedefaultendpunct}{\mcitedefaultseppunct}\relax
\EndOfBibitem
\bibitem[Foffi \emph{et~al.}(2002)Foffi, Dawson, Buldyrev, Sciortino,
  Zaccarelli, and Tartaglia]{foffi2002evidence}
G.~Foffi, K.~A. Dawson, S.~V. Buldyrev, F.~Sciortino, E.~Zaccarelli and
  P.~Tartaglia, \emph{Physical Review E}, 2002, \textbf{65}, 050802\relax
\mciteBstWouldAddEndPuncttrue
\mciteSetBstMidEndSepPunct{\mcitedefaultmidpunct}
{\mcitedefaultendpunct}{\mcitedefaultseppunct}\relax
\EndOfBibitem
\bibitem[Zaccarelli \emph{et~al.}(2003)Zaccarelli, Foffi, Sciortino, and
  Tartaglia]{zaccarelli2003activated}
E.~Zaccarelli, G.~Foffi, F.~Sciortino and P.~Tartaglia, \emph{Physical Review
  Letters}, 2003, \textbf{91}, 108301\relax
\mciteBstWouldAddEndPuncttrue
\mciteSetBstMidEndSepPunct{\mcitedefaultmidpunct}
{\mcitedefaultendpunct}{\mcitedefaultseppunct}\relax
\EndOfBibitem
\bibitem[Sciortino \emph{et~al.}(2003)Sciortino, Tartaglia, and
  Zaccarelli]{sciortino2003evidence}
F.~Sciortino, P.~Tartaglia and E.~Zaccarelli, \emph{Physical Review Letters},
  2003, \textbf{91}, 268301\relax
\mciteBstWouldAddEndPuncttrue
\mciteSetBstMidEndSepPunct{\mcitedefaultmidpunct}
{\mcitedefaultendpunct}{\mcitedefaultseppunct}\relax
\EndOfBibitem
\bibitem[Zaccarelli \emph{et~al.}(2004)Zaccarelli, Sciortino, and
  Tartaglia]{zaccarelli2004numerical}
E.~Zaccarelli, F.~Sciortino and P.~Tartaglia, \emph{Journal of Physics:
  Condensed Matter}, 2004, \textbf{16}, S4849\relax
\mciteBstWouldAddEndPuncttrue
\mciteSetBstMidEndSepPunct{\mcitedefaultmidpunct}
{\mcitedefaultendpunct}{\mcitedefaultseppunct}\relax
\EndOfBibitem
\bibitem[Foffi \emph{et~al.}(2004)Foffi, Sciortino, Zaccarelli, and
  Tartaglia]{foffi2004dynamical}
G.~Foffi, F.~Sciortino, E.~Zaccarelli and P.~Tartaglia, \emph{Journal of
  Physics: Condensed Matter}, 2004, \textbf{16}, S3791\relax
\mciteBstWouldAddEndPuncttrue
\mciteSetBstMidEndSepPunct{\mcitedefaultmidpunct}
{\mcitedefaultendpunct}{\mcitedefaultseppunct}\relax
\EndOfBibitem
\bibitem[Saika-Voivod \emph{et~al.}(2004)Saika-Voivod, Zaccarelli, Sciortino,
  Buldyrev, and Tartaglia]{saika2004effect}
I.~Saika-Voivod, E.~Zaccarelli, F.~Sciortino, S.~V. Buldyrev and P.~Tartaglia,
  \emph{Physical Review E}, 2004, \textbf{70}, 041401\relax
\mciteBstWouldAddEndPuncttrue
\mciteSetBstMidEndSepPunct{\mcitedefaultmidpunct}
{\mcitedefaultendpunct}{\mcitedefaultseppunct}\relax
\EndOfBibitem
\bibitem[Reichman \emph{et~al.}(2005)Reichman, Rabani, and
  Geissler]{reichman2005comparison}
D.~R. Reichman, E.~Rabani and P.~L. Geissler, \emph{The Journal of Physical
  Chemistry B}, 2005, \textbf{109}, 14654--14658\relax
\mciteBstWouldAddEndPuncttrue
\mciteSetBstMidEndSepPunct{\mcitedefaultmidpunct}
{\mcitedefaultendpunct}{\mcitedefaultseppunct}\relax
\EndOfBibitem
\bibitem[Moreno and Colmenero(2006)]{moreno2006anomalous}
A.~J. Moreno and J.~Colmenero, \emph{Physical Review E}, 2006, \textbf{74},
  021409\relax
\mciteBstWouldAddEndPuncttrue
\mciteSetBstMidEndSepPunct{\mcitedefaultmidpunct}
{\mcitedefaultendpunct}{\mcitedefaultseppunct}\relax
\EndOfBibitem
\bibitem[Gonz{\'a}lez-L{\'o}pez \emph{et~al.}(2021)Gonz{\'a}lez-L{\'o}pez,
  Shivam, Zheng, Ciamarra, and Lerner]{gonzalez2021mechanical1}
K.~Gonz{\'a}lez-L{\'o}pez, M.~Shivam, Y.~Zheng, M.~P. Ciamarra and E.~Lerner,
  \emph{Physical Review E}, 2021, \textbf{103}, 022605\relax
\mciteBstWouldAddEndPuncttrue
\mciteSetBstMidEndSepPunct{\mcitedefaultmidpunct}
{\mcitedefaultendpunct}{\mcitedefaultseppunct}\relax
\EndOfBibitem
\bibitem[Gonz{\'a}lez-L{\'o}pez \emph{et~al.}(2021)Gonz{\'a}lez-L{\'o}pez,
  Shivam, Zheng, Ciamarra, and Lerner]{gonzalez2021mechanical2}
K.~Gonz{\'a}lez-L{\'o}pez, M.~Shivam, Y.~Zheng, M.~P. Ciamarra and E.~Lerner,
  \emph{Physical Review E}, 2021, \textbf{103}, 022606\relax
\mciteBstWouldAddEndPuncttrue
\mciteSetBstMidEndSepPunct{\mcitedefaultmidpunct}
{\mcitedefaultendpunct}{\mcitedefaultseppunct}\relax
\EndOfBibitem
\bibitem[Mallamace \emph{et~al.}(2000)Mallamace, Gambadauro, Micali, Tartaglia,
  Liao, and Chen]{mallamace2000kinetic}
F.~Mallamace, P.~Gambadauro, N.~Micali, P.~Tartaglia, C.~Liao and S.-H. Chen,
  \emph{Physical Review Letters}, 2000, \textbf{84}, 5431\relax
\mciteBstWouldAddEndPuncttrue
\mciteSetBstMidEndSepPunct{\mcitedefaultmidpunct}
{\mcitedefaultendpunct}{\mcitedefaultseppunct}\relax
\EndOfBibitem
\bibitem[Eckert and Bartsch(2002)]{eckert2002re}
T.~Eckert and E.~Bartsch, \emph{Physical Review Letters}, 2002, \textbf{89},
  125701\relax
\mciteBstWouldAddEndPuncttrue
\mciteSetBstMidEndSepPunct{\mcitedefaultmidpunct}
{\mcitedefaultendpunct}{\mcitedefaultseppunct}\relax
\EndOfBibitem
\bibitem[Pham \emph{et~al.}(2002)Pham, Puertas, Bergenholtz, Egelhaaf,
  Moussa{\i}d, Pusey, Schofield, Cates, Fuchs, and Poon]{pham2002multiple}
K.~N. Pham, A.~M. Puertas, J.~Bergenholtz, S.~U. Egelhaaf, A.~Moussa{\i}d,
  P.~N. Pusey, A.~B. Schofield, M.~E. Cates, M.~Fuchs and W.~C. Poon,
  \emph{Science}, 2002, \textbf{296}, 104--106\relax
\mciteBstWouldAddEndPuncttrue
\mciteSetBstMidEndSepPunct{\mcitedefaultmidpunct}
{\mcitedefaultendpunct}{\mcitedefaultseppunct}\relax
\EndOfBibitem
\bibitem[Pham \emph{et~al.}(2004)Pham, Egelhaaf, Pusey, and
  Poon]{pham2004glasses}
K.~Pham, S.~Egelhaaf, P.~Pusey and W.~C. Poon, \emph{Physical Review E}, 2004,
  \textbf{69}, 011503\relax
\mciteBstWouldAddEndPuncttrue
\mciteSetBstMidEndSepPunct{\mcitedefaultmidpunct}
{\mcitedefaultendpunct}{\mcitedefaultseppunct}\relax
\EndOfBibitem
\bibitem[Kaufman and Weitz(2006)]{kaufman2006direct}
L.~J. Kaufman and D.~A. Weitz, \emph{The Journal of Chemical Physics}, 2006,
  \textbf{125}, 074716\relax
\mciteBstWouldAddEndPuncttrue
\mciteSetBstMidEndSepPunct{\mcitedefaultmidpunct}
{\mcitedefaultendpunct}{\mcitedefaultseppunct}\relax
\EndOfBibitem
\bibitem[Buzzaccaro \emph{et~al.}(2007)Buzzaccaro, Rusconi, and
  Piazza]{buzzaccaro2007sticky}
S.~Buzzaccaro, R.~Rusconi and R.~Piazza, \emph{Physical Review Letters}, 2007,
  \textbf{99}, 098301\relax
\mciteBstWouldAddEndPuncttrue
\mciteSetBstMidEndSepPunct{\mcitedefaultmidpunct}
{\mcitedefaultendpunct}{\mcitedefaultseppunct}\relax
\EndOfBibitem
\bibitem[Lu \emph{et~al.}(2008)Lu, Zaccarelli, Ciulla, Schofield, Sciortino,
  and Weitz]{lu2008gelation}
P.~J. Lu, E.~Zaccarelli, F.~Ciulla, A.~B. Schofield, F.~Sciortino and D.~A.
  Weitz, \emph{Nature}, 2008, \textbf{453}, 499--503\relax
\mciteBstWouldAddEndPuncttrue
\mciteSetBstMidEndSepPunct{\mcitedefaultmidpunct}
{\mcitedefaultendpunct}{\mcitedefaultseppunct}\relax
\EndOfBibitem
\bibitem[Zhang \emph{et~al.}(2011)Zhang, Yunker, Habdas, and
  Yodh]{zhang2011cooperative}
Z.~Zhang, P.~J. Yunker, P.~Habdas and A.~Yodh, \emph{Physical Review Letters},
  2011, \textbf{107}, 208303\relax
\mciteBstWouldAddEndPuncttrue
\mciteSetBstMidEndSepPunct{\mcitedefaultmidpunct}
{\mcitedefaultendpunct}{\mcitedefaultseppunct}\relax
\EndOfBibitem
\bibitem[Brown \emph{et~al.}(2016)Brown, Iwanicki, Gratale, Ma, Yodh, and
  Habdas]{brown2016correlated}
Z.~Brown, M.~J. Iwanicki, M.~D. Gratale, X.~Ma, A.~Yodh and P.~Habdas,
  \emph{EPL (Europhysics Letters)}, 2016, \textbf{115}, 68003\relax
\mciteBstWouldAddEndPuncttrue
\mciteSetBstMidEndSepPunct{\mcitedefaultmidpunct}
{\mcitedefaultendpunct}{\mcitedefaultseppunct}\relax
\EndOfBibitem
\bibitem[Fullerton and Berthier(2020)]{fullerton2020glassy}
C.~J. Fullerton and L.~Berthier, \emph{Physical Review Letters}, 2020,
  \textbf{125}, 258004\relax
\mciteBstWouldAddEndPuncttrue
\mciteSetBstMidEndSepPunct{\mcitedefaultmidpunct}
{\mcitedefaultendpunct}{\mcitedefaultseppunct}\relax
\EndOfBibitem
\bibitem[Hansen and McDonald(2013)]{hansen1990theory}
J.-P. Hansen and I.~R. McDonald, \emph{Theory of simple liquids}, Elsevier,
  Amsterdam, 2013\relax
\mciteBstWouldAddEndPuncttrue
\mciteSetBstMidEndSepPunct{\mcitedefaultmidpunct}
{\mcitedefaultendpunct}{\mcitedefaultseppunct}\relax
\EndOfBibitem
\bibitem[Ciarella(2021)]{SimonePhdthesis}
S.~Ciarella, \emph{PhD thesis}, Eindhoven University of Technology, 2021\relax
\mciteBstWouldAddEndPuncttrue
\mciteSetBstMidEndSepPunct{\mcitedefaultmidpunct}
{\mcitedefaultendpunct}{\mcitedefaultseppunct}\relax
\EndOfBibitem
\bibitem[Janssen \emph{et~al.}(2016)Janssen, Mayer, and
  Reichman]{janssen2016schematic}
L.~M.~C. Janssen, P.~Mayer and D.~R. Reichman, \emph{Journal of Statistical
  Mechanics: Theory and Experiment}, 2016, \textbf{2016}, 054049\relax
\mciteBstWouldAddEndPuncttrue
\mciteSetBstMidEndSepPunct{\mcitedefaultmidpunct}
{\mcitedefaultendpunct}{\mcitedefaultseppunct}\relax
\EndOfBibitem
\bibitem[Biezemans \emph{et~al.}(2020)Biezemans, Ciarella, {\c{C}}aylak,
  Baumeier, and Janssen]{biezemans2020glassy}
R.~A. Biezemans, S.~Ciarella, O.~{\c{C}}aylak, B.~Baumeier and L.~M.~C.
  Janssen, \emph{Journal of Statistical Mechanics: Theory and Experiment},
  2020,  103301\relax
\mciteBstWouldAddEndPuncttrue
\mciteSetBstMidEndSepPunct{\mcitedefaultmidpunct}
{\mcitedefaultendpunct}{\mcitedefaultseppunct}\relax
\EndOfBibitem
\bibitem[Meyer(2000)]{meyer2000matrix}
C.~D. Meyer, \emph{Matrix analysis and applied linear algebra}, Siam, 2000,
  vol.~71\relax
\mciteBstWouldAddEndPuncttrue
\mciteSetBstMidEndSepPunct{\mcitedefaultmidpunct}
{\mcitedefaultendpunct}{\mcitedefaultseppunct}\relax
\EndOfBibitem
\bibitem[Arnol'd(2003)]{arnol2003catastrophe}
V.~I. Arnol'd, \emph{Catastrophe theory}, Springer Science \& Business Media,
  2003\relax
\mciteBstWouldAddEndPuncttrue
\mciteSetBstMidEndSepPunct{\mcitedefaultmidpunct}
{\mcitedefaultendpunct}{\mcitedefaultseppunct}\relax
\EndOfBibitem
\bibitem[Dawson \emph{et~al.}(2000)Dawson, Foffi, Fuchs, G{\"o}tze, Sciortino,
  Sperl, Tartaglia, Voigtmann, and Zaccarelli]{dawson2000higher}
K.~Dawson, G.~Foffi, M.~Fuchs, W.~G{\"o}tze, F.~Sciortino, M.~Sperl,
  P.~Tartaglia, T.~Voigtmann and E.~Zaccarelli, \emph{Physical Review E}, 2000,
  \textbf{63}, 011401\relax
\mciteBstWouldAddEndPuncttrue
\mciteSetBstMidEndSepPunct{\mcitedefaultmidpunct}
{\mcitedefaultendpunct}{\mcitedefaultseppunct}\relax
\EndOfBibitem
\bibitem[Gotze and Sjogren(1989)]{gotze1989logarithmic}
W.~Gotze and L.~Sjogren, \emph{Journal of Physics: Condensed Matter}, 1989,
  \textbf{1}, 4203\relax
\mciteBstWouldAddEndPuncttrue
\mciteSetBstMidEndSepPunct{\mcitedefaultmidpunct}
{\mcitedefaultendpunct}{\mcitedefaultseppunct}\relax
\EndOfBibitem
\bibitem[G{\"o}tze and Sperl(2002)]{gotze2002logarithmic}
W.~G{\"o}tze and M.~Sperl, \emph{Physical Review E}, 2002, \textbf{66},
  011405\relax
\mciteBstWouldAddEndPuncttrue
\mciteSetBstMidEndSepPunct{\mcitedefaultmidpunct}
{\mcitedefaultendpunct}{\mcitedefaultseppunct}\relax
\EndOfBibitem
\bibitem[Franosch \emph{et~al.}(1997)Franosch, Fuchs, G{\"o}tze, Mayr, and
  Singh]{franosch1997asymptotic}
T.~Franosch, M.~Fuchs, W.~G{\"o}tze, M.~R. Mayr and A.~Singh, \emph{Physical
  Review E}, 1997, \textbf{55}, 7153\relax
\mciteBstWouldAddEndPuncttrue
\mciteSetBstMidEndSepPunct{\mcitedefaultmidpunct}
{\mcitedefaultendpunct}{\mcitedefaultseppunct}\relax
\EndOfBibitem
\bibitem[Fuchs \emph{et~al.}(1991)Fuchs, Gotze, Hofacker, and
  Latz]{fuchs1991comments}
M.~Fuchs, W.~Gotze, I.~Hofacker and A.~Latz, \emph{Journal of Physics:
  Condensed Matter}, 1991, \textbf{3}, 5047\relax
\mciteBstWouldAddEndPuncttrue
\mciteSetBstMidEndSepPunct{\mcitedefaultmidpunct}
{\mcitedefaultendpunct}{\mcitedefaultseppunct}\relax
\EndOfBibitem
\bibitem[G{\"o}tze and Sperl(2003)]{gotze2003higher}
W.~G{\"o}tze and M.~Sperl, \emph{Journal of Physics: Condensed Matter}, 2003,
  \textbf{15}, S869\relax
\mciteBstWouldAddEndPuncttrue
\mciteSetBstMidEndSepPunct{\mcitedefaultmidpunct}
{\mcitedefaultendpunct}{\mcitedefaultseppunct}\relax
\EndOfBibitem
\bibitem[Fabbian \emph{et~al.}(1999)Fabbian, G{\"o}tze, Sciortino, Tartaglia,
  and Thiery]{fabbian1999ideal}
L.~Fabbian, W.~G{\"o}tze, F.~Sciortino, P.~Tartaglia and F.~Thiery,
  \emph{Physical Review E}, 1999, \textbf{59}, R1347\relax
\mciteBstWouldAddEndPuncttrue
\mciteSetBstMidEndSepPunct{\mcitedefaultmidpunct}
{\mcitedefaultendpunct}{\mcitedefaultseppunct}\relax
\EndOfBibitem
\bibitem[Bergenholtz and Fuchs(1999)]{bergenholtz1999nonergodicity}
J.~Bergenholtz and M.~Fuchs, \emph{Physical Review E}, 1999, \textbf{59},
  5706\relax
\mciteBstWouldAddEndPuncttrue
\mciteSetBstMidEndSepPunct{\mcitedefaultmidpunct}
{\mcitedefaultendpunct}{\mcitedefaultseppunct}\relax
\EndOfBibitem
\bibitem[Ciarella \emph{et~al.}(2021)Ciarella, Luo, Debets, and
  Janssen]{ciarella2021multi}
S.~Ciarella, C.~Luo, V.~E. Debets and L.~Janssen, \emph{arXiv preprint
  arXiv:2103.16522}, 2021\relax
\mciteBstWouldAddEndPuncttrue
\mciteSetBstMidEndSepPunct{\mcitedefaultmidpunct}
{\mcitedefaultendpunct}{\mcitedefaultseppunct}\relax
\EndOfBibitem
\bibitem[Nandi \emph{et~al.}(2014)Nandi, Biroli, Bouchaud, Miyazaki, and
  Reichman]{nandi2014critical}
S.~K. Nandi, G.~Biroli, J.-P. Bouchaud, K.~Miyazaki and D.~R. Reichman,
  \emph{Physical Review Letters}, 2014, \textbf{113}, 245701\relax
\mciteBstWouldAddEndPuncttrue
\mciteSetBstMidEndSepPunct{\mcitedefaultmidpunct}
{\mcitedefaultendpunct}{\mcitedefaultseppunct}\relax
\EndOfBibitem
\bibitem[Robinson \emph{et~al.}(2019)Robinson, Turci, Roth, and
  Royall]{robinson2019morphometric}
J.~F. Robinson, F.~Turci, R.~Roth and C.~P. Royall, \emph{Physical Review
  Letters}, 2019, \textbf{122}, 068004\relax
\mciteBstWouldAddEndPuncttrue
\mciteSetBstMidEndSepPunct{\mcitedefaultmidpunct}
{\mcitedefaultendpunct}{\mcitedefaultseppunct}\relax
\EndOfBibitem
\bibitem[Scopigno \emph{et~al.}(2003)Scopigno, Ruocco, Sette, and
  Monaco]{scopigno2003fragility}
T.~Scopigno, G.~Ruocco, F.~Sette and G.~Monaco, \emph{Science}, 2003,
  \textbf{302}, 849--852\relax
\mciteBstWouldAddEndPuncttrue
\mciteSetBstMidEndSepPunct{\mcitedefaultmidpunct}
{\mcitedefaultendpunct}{\mcitedefaultseppunct}\relax
\EndOfBibitem
\bibitem[Martinez and Angell(2001)]{martinez2001thermodynamic}
L.-M. Martinez and C.~Angell, \emph{Nature}, 2001, \textbf{410}, 663--667\relax
\mciteBstWouldAddEndPuncttrue
\mciteSetBstMidEndSepPunct{\mcitedefaultmidpunct}
{\mcitedefaultendpunct}{\mcitedefaultseppunct}\relax
\EndOfBibitem
\bibitem[Sciortino and Kob(2001)]{sciortino2001debye}
F.~Sciortino and W.~Kob, \emph{Physical Review Letters}, 2001, \textbf{86},
  648\relax
\mciteBstWouldAddEndPuncttrue
\mciteSetBstMidEndSepPunct{\mcitedefaultmidpunct}
{\mcitedefaultendpunct}{\mcitedefaultseppunct}\relax
\EndOfBibitem
\end{mcitethebibliography}
\bibliographystyle{sticky_hard_sphere} 

\end{document}